\newcounter{myctr}

\documentclass[10pt,letterpaper]{article}
\usepackage[top=0.85in,footskip=0.75in,marginparwidth=2in]{geometry}

\usepackage{url}
\usepackage{nth}
\usepackage{float}
\usepackage{array,multirow}
\usepackage[table, dvipsnames]{xcolor}

\usepackage{enumerate}

\usepackage[T1]{fontenc}
\usepackage{makeidx}
\usepackage[normalem]{ulem}
\usepackage{color}
\usepackage{bm}% bold math
\usepackage{hyperref}   % use for hypertext links, including those to external documents and URLs
\usepackage{tabularx} % stretched tables
\usepackage{wrapfig}
\usepackage{setspace}
\usepackage{subfigure}
\usepackage{nameref,hyperref}
\usepackage{url}
\usepackage{nth}
\usepackage{float}
\usepackage{array,multirow}
\usepackage{bm}% bold math
\usepackage{hyperref}
\usepackage[right]{lineno}

\usepackage{microtype}
\DisableLigatures[f]{encoding = *, family = * }

\usepackage{changepage}

\usepackage[aboveskip=1pt,labelfont=bf,labelsep=period,singlelinecheck=off]{caption}

\makeatletter
\renewcommand{\@biblabel}[1]{\quad#1.}
\makeatother

\usepackage{lastpage,fancyhdr,graphicx}
\usepackage{epstopdf}
\fancyheadoffset[L]{2.25in}
\fancyfootoffset[L]{2.25in}

\usepackage{color}

\definecolor{Gray}{gray}{.25}

\usepackage{graphicx}

\usepackage{sidecap}

\usepackage{wrapfig}
\usepackage[pscoord]{eso-pic}
\usepackage[fulladjust]{marginnote}
\reversemarginpar

\begin{document}

\makeatletter
\def\@biblabel#1{[#1]}
\makeatother

\begin{flushleft}
{\Large
\textbf\newline{Quantifying the relation between performance and success \\in soccer}
}
\newline
% authors go here:
\\
Luca Pappalardo\textsuperscript{1,2},
Paolo Cintia\textsuperscript{2}
\\
\bigskip
\bf{1} Department of Computer Science, University of Pisa, Italy
\\
\bf{2} ISTI-CNR, Pisa, Italy
\\
\bigskip
lpappalardo@di.unipi.it, paolo.cintia@isti.cnr.it

\end{flushleft}

\begin{abstract}
The availability of massive data about sports activities offers nowadays the opportunity to quantify the relation between performance and success. In this study, we analyze more than 6,000 games and 10 million events in six European leagues and investigate this relation in soccer competitions. We discover that a team's position in a competition's final ranking is significantly related to its typical performance, as described by a set of technical features extracted from the soccer data. Moreover we find that, while victory and defeats can be explained by the team's performance during a game, it is difficult to detect draws by using a machine learning approach. We then simulate the outcomes of an entire season of each league only relying on technical data, i.e. excluding the goals scored, exploiting a machine learning model trained on data from past seasons. The simulation produces a team ranking (the PC ranking) which is close to the actual ranking, suggesting that a complex systems' view on soccer has the potential of revealing hidden patterns regarding the relation between performance and success.
\end{abstract}

\section{Introduction}
\label{intro}
Soccer is the world's most popular sports and keeps attracting more and more fans and investors \cite{dobson2011}. Given the nature of the game itself, where two teams of eleven players produce a huge number of interactions, the statistical analysis of soccer games have fascinated scientists, coaches, and experts. Already in the 1950s, Charles Reep collected soccer statistics by hand performing the first statistical analysis of soccer \cite{reep1,reep2}, while in the 1970s coach Valeriy Lobanovskyi defined schemes and tactics by using one of the first prototypes of a computer and the help of a statistician \cite{anderson2013numbers}. Apart from these seminal initiatives, academic work on soccer analytics has been deterred for decades by the limited availability of detailed data.

Nowadays, the data revolution has the potential to rapidly change this scenario, thanks to new sensing technologies that provide high-fidelity data streams extracted from every game, such as the spatio-temporal trajectories of players \cite{DBLP:journals/corr/GudmundssonH16,rossi2017effective,rossi2017who} and all the events that occur on the field \cite{stein2017how,cintia2015network,harsh}. Recently, several studies relied on these data to propose metrics which quantify specific aspects of soccer performance \cite{clemente2015using,horton2014classification,taki2000visualization,kang,lucey2014quality,pena2012network,brooks2016developing}. However, a quantification of the relationships between technical performance and success in soccer is still missing in the literature. Technical performance measures how teams behave on the field, while success measures the teams' achievement during a competition, such as a game outcome or the position in the final ranking. While it is common knowledge that the most successful teams tend to score more goals than the opponents \cite{harsh,anderson2013numbers,heuer2009fitness,heuer2013perfekte}, it is not clear what the contribution is of other measurable aspects (e.g., passes, shots, etc.) and to what extent \emph{their combination} can explain a team's success. The state of the art mainly focuses on specific aspects of a team's behavior (typically passes and shots) \cite{brooks2016developing,harsh,lucey2014quality}, while a multidimensional view of soccer performance, seen as a combination of different technical features, has not been discussed in the scientific literature. Which technical features influence a team's success and how do they play in combination? Which characteristics affect a game's outcome? Can we construct a ranking for soccer teams which relies solely on their technical behavior?

To answer these questions, we analyze massive soccer logs describing all games in six European soccer leagues during three seasons -- 145 teams, 6,396 games, and 10 million game events. We first define a team's technical performance as a multidimensional vector of features extracted from soccer logs. These features cover several aspects of a team's behavior such as goalkeeping, intercepts, tackles, dribbles, passes, shots and fouls. We then develop our analysis in three different directions. 

First, we quantify the relation between technical performance and success, as measured by the teams' number of points obtained at the end of a competition \cite{bennaim2006,harsh}. 
We demonstrate that technical performance can explain more than half of the variance in a team's final ranking and investigate the importance of every technical feature to a team's success. Second, we use machine learning to train a game outcome predictor and understand to what extent a team's technical performance can explain a game outcome. We demonstrate that, while victories and defeats can be explained by technical performance, it is difficult to detect draws. 
Finally, we exploit the game outcome predictor to conduct an experiment consisting of a complete simulation of the six national tournaments -- Premier League, Serie A, La Liga, Bundesliga, Eredivisie, Ligue 1. The outcome of each game is replaced by a synthetic outcome (victory, defeat or draw) based on a game outcome predictor trained on the previous seasons. For each competition, the simulation produces a team ranking (the PC ranking) entirely based on the quantified relation between technical performance and game outcome. Despite the low accuracy in detecting draws, in the long run game predictor generates rankings that are similar to the actual ones, highlighting its ability in capturing the complex relationship between performance and success. Our approach opens interesting perspectives to understand to what extent a team's success in a competition can be described by its technical behavior on the field.

\section{Soccer Logs}
\label{sec:data}
We use soccer logs describing all the games in six major soccer leagues in seasons 2013/2014, 2014/2015 and 2015/2016: Premier League, Serie A, La Liga, Bundesliga, Eredivisie, Ligue 1. Although the number of participating clubs varies across the leagues, the same round-robin format is used (\ref{app:round_robin}). Premier League, Serie A, La Liga and Ligue 1 have 20 clubs and 380 games per season each; Bundesliga and Eredivisie have 18 clubs and 306 games per season each. In total, the dataset stores information about 145 clubs and 6,396 games (see Table \ref{tab:data_summary}). 

\begin{figure}
\begin{center}
\includegraphics[scale=0.44]{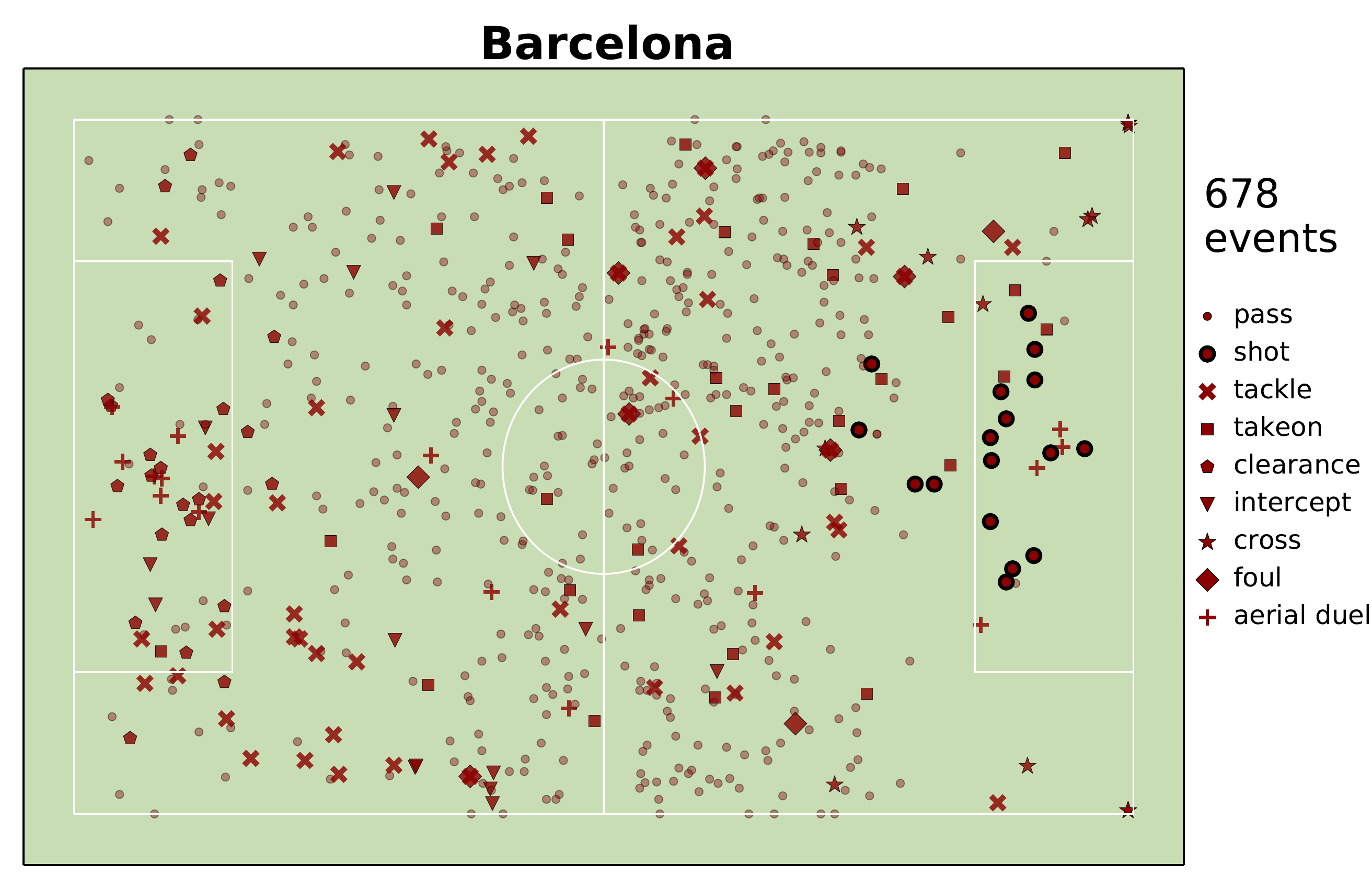}
\includegraphics[scale=0.44]{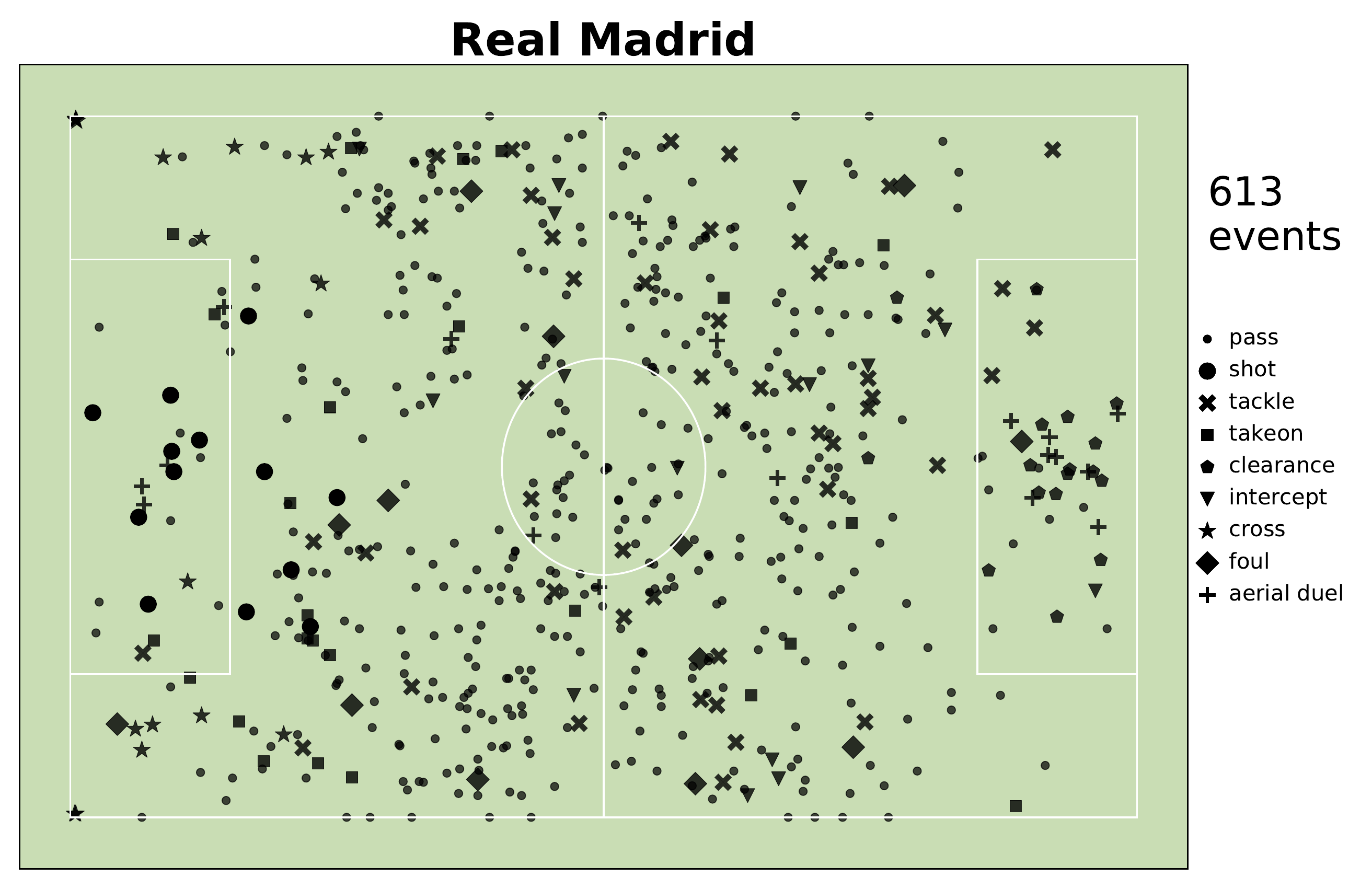}
\caption{\textbf{A visualization of the complexity of a soccer game.} The figures show the positions on the pitch of the events that occurred during the game FC Barcelona vs Real Madrid (Spanish League, \nth{22} March 2015). Every point is plotted on the position of the pitch where the corresponding event was generated. An event can be one of the following types: pass, cross, shot, tackle, dribble, clearance, intercept, foul, aerial duel. We do not show goalkeeping events because the position on the pitch is not available for this event type.}
\label{fig:all_data}
\end{center}
\end{figure}

Every game is described by a sequence of events that occurred on the football pitch, with a total of around 10 million events. Each event consists of a timestamp, the player who generated the event and the position on the field (Table \ref{tab:data}). There are several event types, each corresponding to a type of action a player can perform during a game: passes, crosses, shots, tackles, dribbles, clearances, goalkeeping actions, fouls, intercepts, aerial duels, goals scored and goals conceded.
Table \ref{tab:data} shows a sample of events that occurred during a game in La Liga between FC Barcelona and Real Madrid held on \nth{22} March 2015. Since the size of the pitch is slightly variable from stadium to stadium -- UEFA establishes that it can be 100 to 105 meters long, and 64 to 68 meters wide\footnote{\url{https://en.wikipedia.org/wiki/UEFA_stadium_categories}} -- we normalize both the pitch coordinates in the range $[0, 100]$. The colored row in Table \ref{tab:data} shows an event where player Messi (FC Barcelona) makes a pass from position $(78.3, 40.2)$ of the pitch, 1,389 seconds into the game. 
Figure \ref{fig:all_data} is a visualization of the complexity of the entire game FC Barcelona vs Real Madrid: all the events that occurred on the pitch are plotted on the position where they were generated, with a different marker for every event type. Although it is just a plain visualization of the data without any numerical insight, it immediately reveals the confrontation of two competing teams and the ability of soccer logs to capture a game's complexity. \ref{app:distr} presents some descriptive statistics of the soccer dataset.

\begin{table}[htb]\centering
\def\arraystretch{1.2}
\begin{tabular}{|>{\columncolor[gray]{0.95}}c|r|p{7.5cm}|}
\hline
\textbf{seasons} & 3 & 2013/2014, 2014/2015 and 2015/2016\\
\hline
\textbf{leagues} & 6 & Premier League, Ligue 1, Bundesliga, Serie A, La Liga, Eredivisie\\
\hline
\textbf{teams} & 145 & 
\begin{tabular}{c c}
20 & Premier League, Ligue1, Serie A, La Liga\\
18 & Bundesliga, Eredivisie\\
\end{tabular}
\\
\hline
\textbf{games} & 6,396 & 
\begin{tabular}{c c}
380 & Premier League, Ligue1, Serie A, La Liga\\
306 & Bundesliga, Eredivisie\\
\end{tabular}\\
\hline
\textbf{events} & $\approx$ $10^7$ & around 1,600 events per game\\
\hline
\end{tabular}
\caption{\textbf{Description of the soccer dataset.} It covers 3 seasons, 6 leagues and 145 different clubs, with 6,396 games described by around 10 million events.}
\label{tab:data_summary}
\end{table}

\begin{table}\centering
\begin{tabular}{|c|c|c|c|c|}
\hline
\textbf{team} & \textbf{player} & \textbf{event} & \textbf{pos} & \textbf{time}\\
\hline
\vdots &\vdots &\vdots &\vdots &\vdots\\ 
Barcelona & Iniesta & pass & (68.9,55) & 1253\\
Barcelona & Messi & shot & (88.4,60.3) & 1343\\
\cellcolor{blue!20}Barcelona & \cellcolor{blue!20}Messi & \cellcolor{blue!20}pass &\cellcolor{blue!20}(78.3,40.2) & \cellcolor{blue!20}1389\\
Barcelona & Alba & shot & (64.4,81.9) & 1390\\
R Madrid & Carvajal & pass & (46.3,17.8) & 1406\\
Barcelona & Mathieu & pass & (40.5,89.9) & 1408\\

Barcelona & Neymar & tackle & (59.5,86.1)  & 1409\\
R Madrid & Carvajal & tackle & (39.2,29.2)  & 1412\\
R Madrid & Carvajal & pass & (46.3,42.3) & 1414\\
R Madrid & Isco & pass & (62.8,67.5) & 1416\\
R Madrid & Ronaldo & pass & (82.8,87.7) & 1422\\
\vdots & \vdots &\vdots &\vdots &\vdots \\
\hline
\end{tabular}
\caption{\textbf{Example of events in the soccer logs.} The events occurred in the game FC Barcelona vs Real Madrid (La Liga, \nth{22} March 2015). Each row is an event and has the following fields: (1) the team who generated the event; (2) the player who generated the event; (3) the type of event (pass, shot, etc.); (4) the position of the event; (5) the timestamp of the event (in seconds since the beginning of the game). The blue row in the table indicates a pass made by player Messi at position $(78.3, 40.2)$ 1,389 seconds into the game. }
\label{tab:data}
\end{table}

\newcommand{\vect}[1]{\boldsymbol{#1}}
\section{Multidimensional performance in soccer}
\label{sec:indicators}
Soccer performance is a multidimensional concept: many features determine team's performance during a game, and each feature describes a different aspect of team's collective behavior. Formally, we describe the performance of a team $A$ in game $g$ as a $n$-dimensional feature vector: 
$$\vect{h}_A^{(g)} = [x_1^{(g)}(A), \dots, x_n^{(g)}(A)],$$ 
where $x_i^{(g)}(A)$ is a feature describing a specific aspect of $A$'s performance in game $g$ for features $i = 1, \dots, n$. The performance feature $x_i^{(g)}(A)$ is computed as the sum of the corresponding performance feature of the players composing the team: $$x_i^{(g)}(A) = \sum_{j = 1}^k x^{(g)}_{i}(j),$$ where $x^{(g)}_i(j)$ indicates the value of performance feature $x_i$ produced by player $j$ of team $A$ in game $g$.\footnote{We tried different aggregations of the players' features, such as average and median. They produce results that are similar to those presented in the paper.} In this study, we consider $n=10$ features, corresponding to the event types in the soccer logs: the number of passes, crosses, shots, tackles, dribbles, clearances, goalkeeping actions, fouls, intercepts and aerial duels generated by the players during a game. This vector can be easily extended by including an arbitrary number of features, built over the standard features we use. For example, we also extract information regarding a team's playing quality: pass precision, dribble precision, tackle precision, cross precision, and a coefficient of the team's attack/defense attitude and a team's spatial and temporal dominance, that includes average team position, speed and accelerations (see \ref{app:quality_features}). We do not exploit these features in the paper since we find that, though individually correlated with a team's success, they do not lead to significantly better predictions (see \ref{app:quality_features}). This confirms existing results in the literature which argue that incorporating more features do not necessarily lead to better predictions, due to the highly unpredictable nature of soccer games \cite{heuer2014optimizing,heuer2013perfekte}. Also, we do not include the goals scored, the goals conceded and goal difference. Including these features would produce trivial correlations preventing us from understanding the impact of technical features during a game or competition. In this study, we consider goals as a first evidence of a team's success rather than a measure of performance: a positive goal difference corresponds to a victory (success) and a negative goal difference to a defeat (failure). 

\begin{figure}[htb!]\centering
\subfigure[]{\includegraphics[scale=0.33]{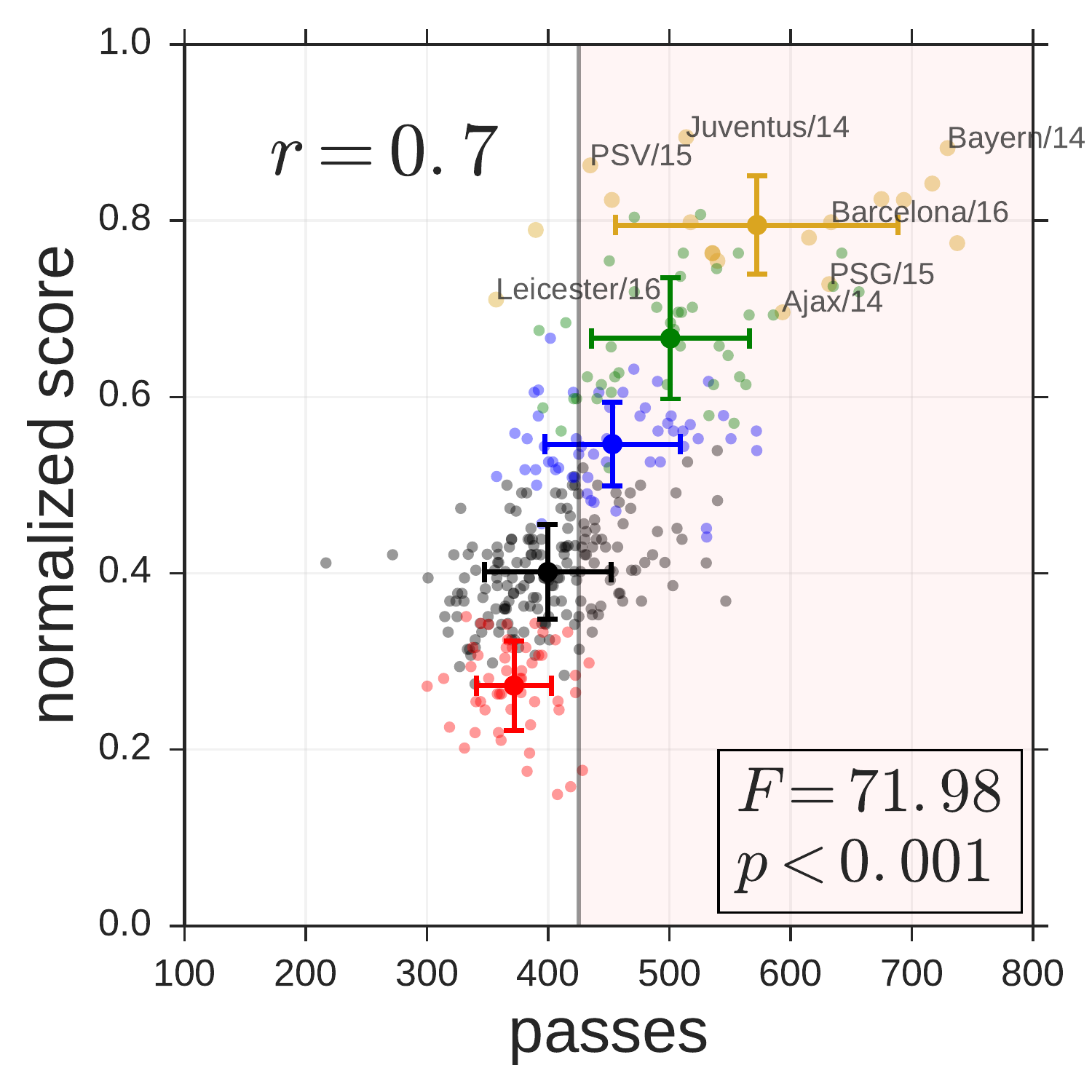}}
\vspace{-2\baselineskip}
\subfigure[]{\includegraphics[scale=0.33]{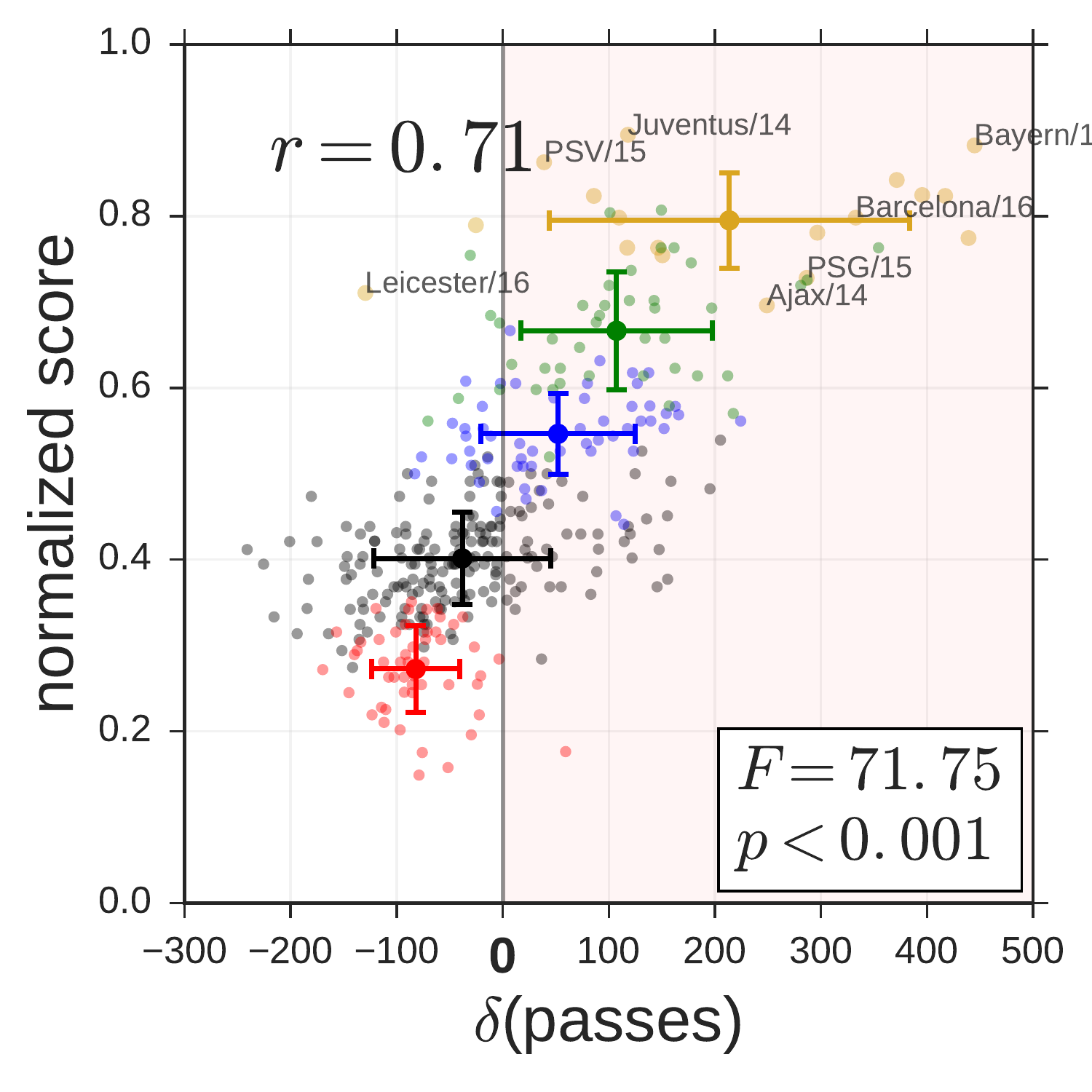}}
\subfigure[]{\includegraphics[scale=0.33]{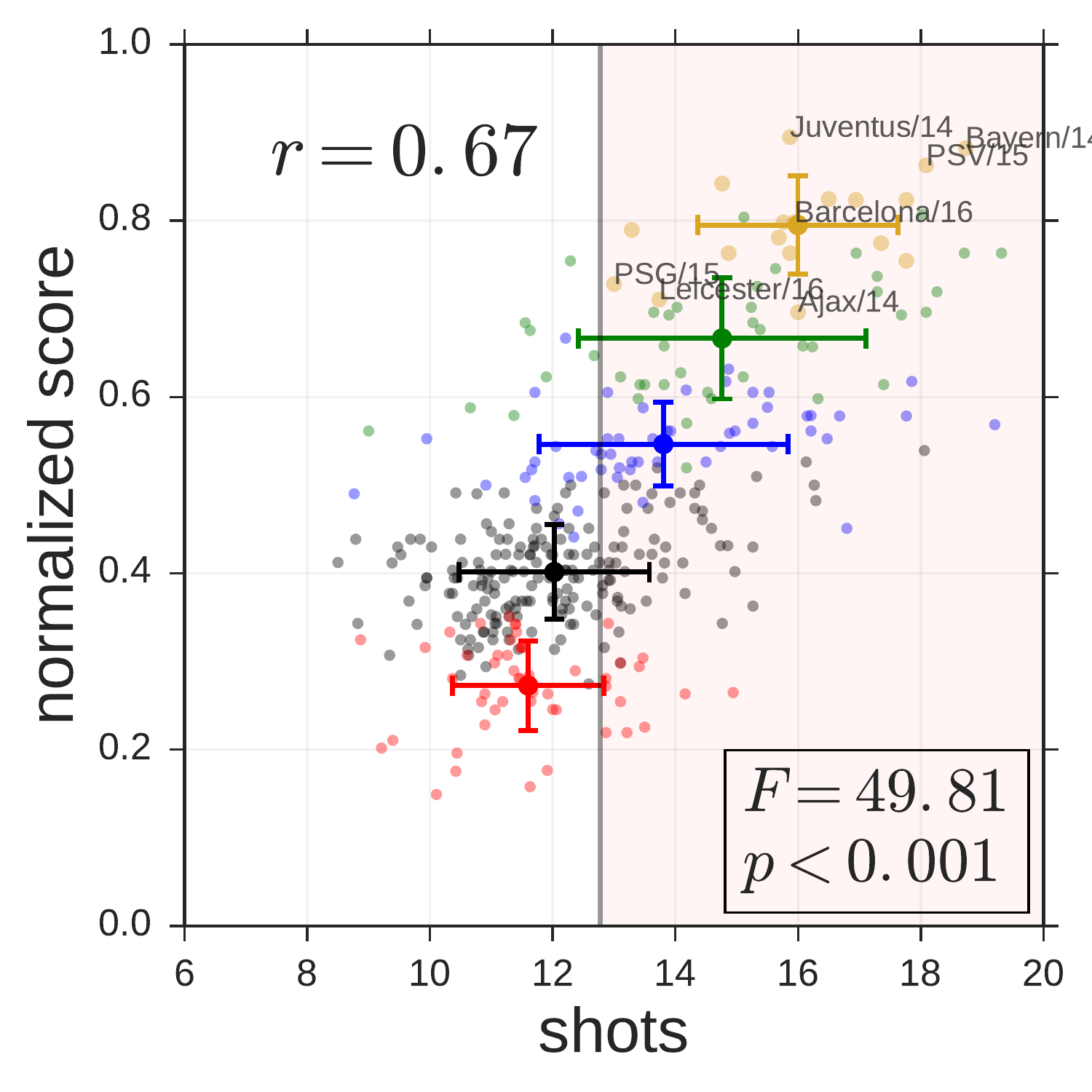}}
\vspace{-2\baselineskip}
\subfigure[]{\includegraphics[scale=0.33]{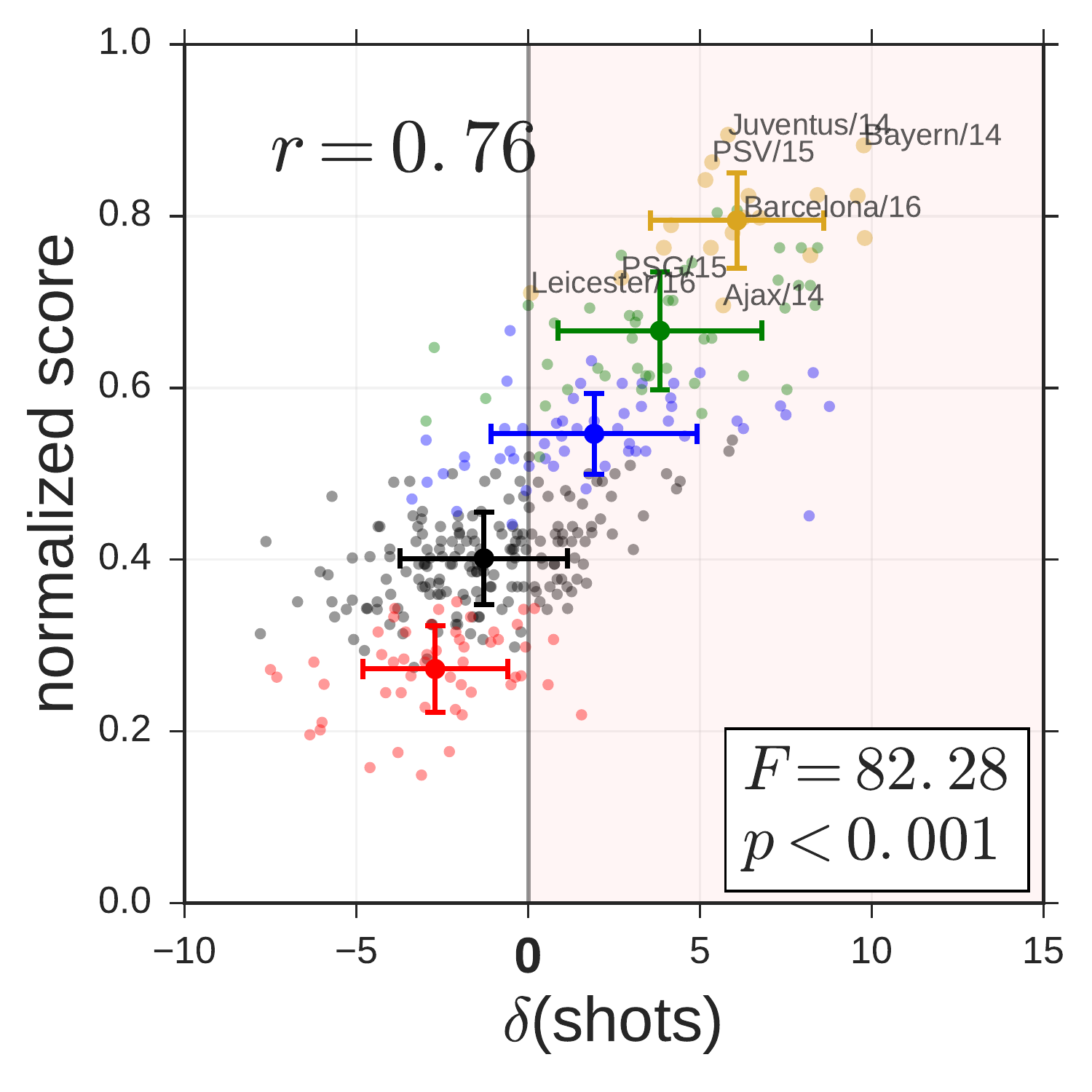}}
\addtocounter{subfigure}{-4}
\subfigure[]{\includegraphics[scale=0.33]{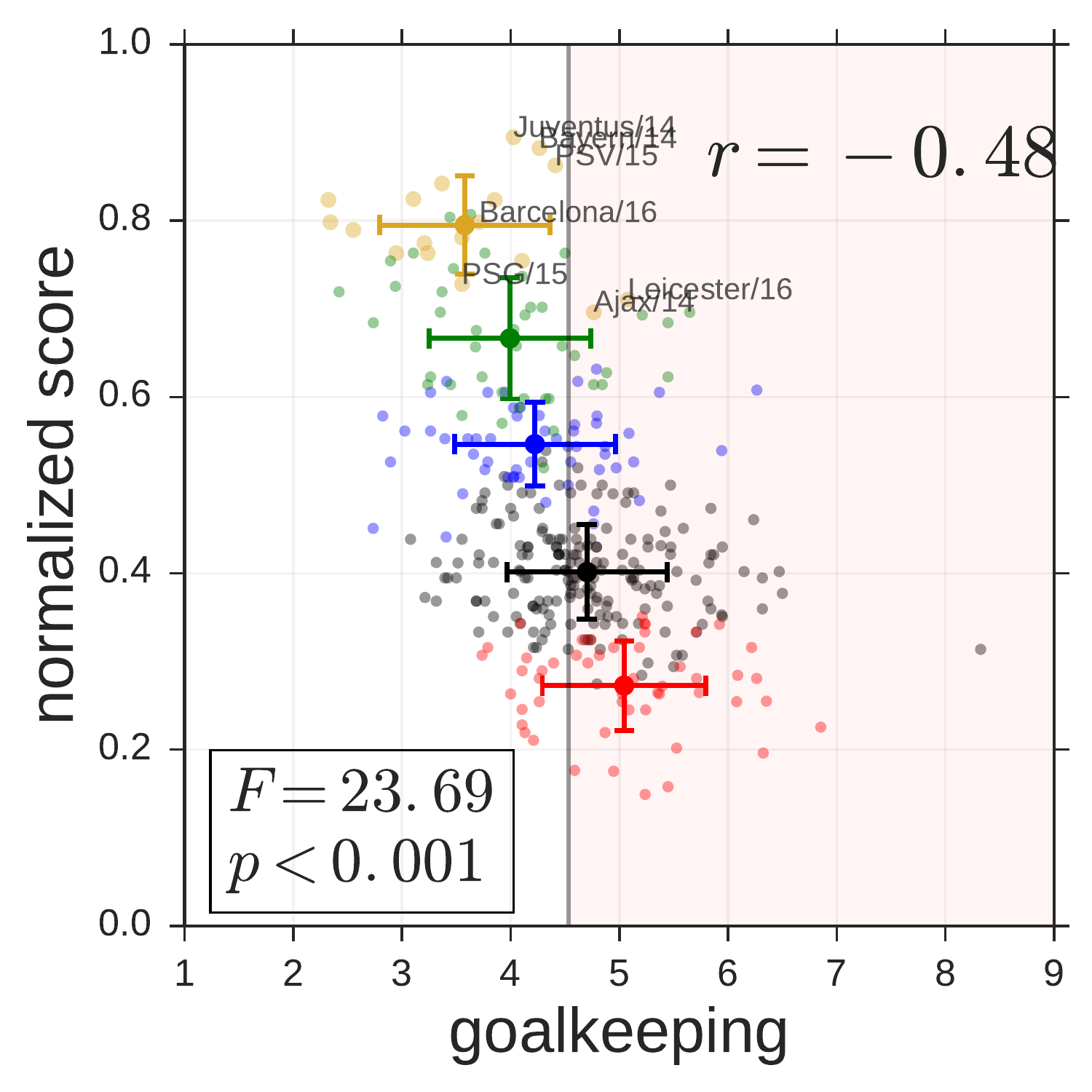}}
\subfigure[]{\includegraphics[scale=0.33]{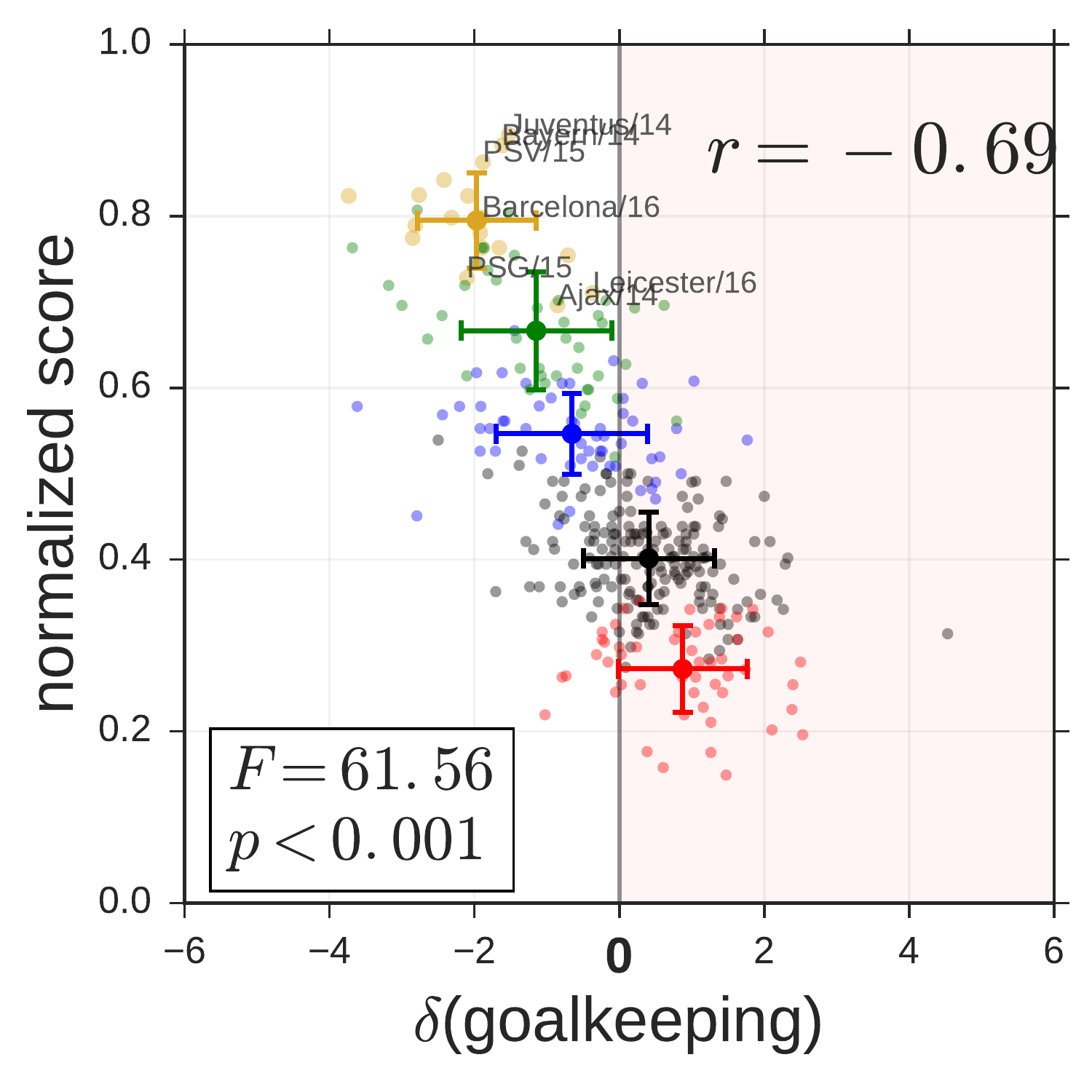}}
\caption{\textbf{Correlation between performance features and success}, for (a) absolute and (b) relative features. Each point is a team in a season. We split the teams in groups according to the final ranking: teams relegated in second division (red), teams in the middle (black), teams in Europa League (blue), teams in Champions League (green) and winners (golden). In the box we indicate F-test statistic and p-value resulting from a one-way ANOVA test to assess whether the features values of two groups differ significantly.}
\label{fig:correlations}
\end{figure}

We use a team's final score at the end of a season to measure its success in a competition. The final score is the number of points gained by a team during a competition, where a team gains 3 points for a victory, 1 point for a draw and no points for a defeat. To account for the different number of clubs in the six leagues, we normalize the final scores in the range $[0, 1]$ (\ref{sec:norm_scores}). Figure \ref{fig:correlations}a-c correlates a team's final score with its \emph{typical absolute performance}, i.e., the feature vector consisting of the average value of the features across all the games of a season: $$\overline{\vect{h}}_A^{(g)} = [\overline{x}_1^{(g)}(A), \dots, \overline{x}_n^{(g)}(A)],$$ where $\overline{x}_i(A) = \frac{1}{N} \sum_{g=1}^N x_i^{(g)}(A)$ and $N$ is the number of games played by team $A$ in the season. We observe a significant Pearson correlation (i.e., $|r|{>}0.1$, $p$-value $< 0.05$) for all the 10 absolute performance features. We find that the average number of passes is slightly more correlated with the final score  ($r{=}0.70$, Figure \ref{fig:correlations}a) than the average number of shots ($r{=}0.67$, Figure \ref{fig:correlations}c). A possible interpretation is that, while the average number of passes is a rather objective proxy for a team's dominance \cite{harsh}, the ratio to which shots convert into goals is low \cite{reep1} since a shot's outcome strongly depends on several conditions, like distance, shooting angle and the quality of the opponent's goalkeeper.
The number of goalkeeping actions, that is a proxy of a team's defensive effort, is negatively correlated with success. 

In Figure \ref{fig:correlations}, we split the teams into different groups according to their position in the ranking at the end of a season: teams relegated in the second division (red), teams in the middle of the ranking (black), teams promoted to Europa League (blue), teams promoted to Champions League (green), and the winners of the national tournament (golden). Based on this grouping, we perform a Tukey's range test \cite{tukey} to determine whether or not there is a significant difference between the teams' feature values in the defined groups. 
We find a statistical difference between the groups for 8 out of 10 features (\ref{app:anova}). For instance, the winners' typical number of passes is significantly different from the typical number of passes of teams in all the other groups. The distributions of goalkeeping actions are significantly different for all the pairs but the pairs involving the two top groups, while the distributions of shots are significantly different for all the pairs but the pairs involving the two bottom groups. The distributions of tackles and crosses show significant difference for none of the groups (\ref{app:anova}).

Both the significance of correlations and the results of the Tukey's test suggest that typical performance features, taken individually, are related to success. 
Top teams and bottom teams behave differently on the field: when the two bottom groups have a value for a feature typically below the average, the top groups have values typically above the average, and vice versa (see Figure \ref{fig:correlations}a). The three top groups, consisting of the most successful teams promoted to continental competitions, perform in a peculiar way.

Figure \ref{fig:correlations}b correlates a team's final score with its \emph{typical relative performance}: 
$$\overline{\vect{r}}_A^{(g)} = [\overline{\delta}_1^{(g)}(A), \dots, \overline{\delta}_n^{(g)}(A)],$$ where $\overline{\delta_i}(A) = \frac{1}{N} \sum_{g=1}^N (x_i^{(g)}(A) {-} x_i^{(g)}(B))$ and $x_i^{(g)}(B)$ is the absolute performance of team $A$'s opponent in game $g$. Relative performance is the difference between the absolute performances of two opposing teams and so relates a team's behavior to the opponent's one \cite{burcu}. We find that the correlations significantly increase using relative performance features, with the exception of passes, clearances and fouls for which the correlation remains approximately the same (Table \ref{tab:pred_results}). This suggests that, for some aspects like passes, the relation between a feature and success relies mainly on a team's specific playing style, regardless of the opponent's playing style. Conversely, for other aspects (e.g., shots) it is the relation between a team's behavior and the opponent's behavior which matters. For the relative performance features, we find a statistical difference between the groups of success for all the features (see \ref{app:anova}). In particular, the Tukey's test reveals that: \emph{(i)} all the relative performance features can discriminate between at least two groups of success; 
\emph{(ii)} the relative performance features can discriminate between more groups of success than the corresponding absolute performance features (\ref{app:anova}).

\begin{table}[htb!]\centering
\begin{tabular}{|m{6cm} |m{6cm}|}
\multicolumn{2}{c}{\large Summary of results -- Section \ref{sec:indicators}}\\
\hline
\multicolumn{1}{|c|}{\large \textbf{Question}} & \multicolumn{1}{c|}{\large Answer}\\
\hline
\centering \textbf{What are absolute and relative performance in soccer?} & Absolute performance is a multidimensional vector where each element indicates a team's technical feature during a game. In relative performance, we relate a team's performance to the performance of the opponent. \\
\hline
\centering \textbf{Are typical absolute performance features correlated to success?} & Yes they are. The typical number of passes, shots and goalkeeping actions are the most correlated.\\
\hline
\centering \textbf{What about the relative performance features?} & They show a higher correlation with success than the absolute performance features. Only passes, clearances and fouls show a similar correlation.\\
\hline
\centering \textbf{Can absolute and relative performance features explain different groups of success?} & Yes, the groups of success show significant differences in terms of both the absolute and the relative performance features.\\
\hline
\end{tabular}
\end{table}

%%%%%%%%%%%%%%%%%%%%%%%
\section{Multidimensional performance and success in competitions}
\label{sec:tournament_success}
To further investigate the relation between performance and success we use ordinary least squares (OLS) \cite{cohen2013applied} to find a linear fit $y = a \overline{x}_i + b$ of each typical absolute performance feature $\overline{x}_i$ to the final score $y$.
We observe a significant coefficient of determination (i.e., $R^2 {>} 0.2$) for three features (see Table \ref{tab:pred_results}): passes ($R^2{=}0.45$), shots ($R^2{=}0.36$) and goalkeeping actions ($R^2{=}0.22$). This indicates that the most descriptive feature can explain up to 45\% of the variance in success. Therefore, no individual feature can fully explain success in soccer, indicating that typical performance can drive success through a combination of performance features. 

\begin{figure}
\subfigure[]{\includegraphics[scale=0.4]{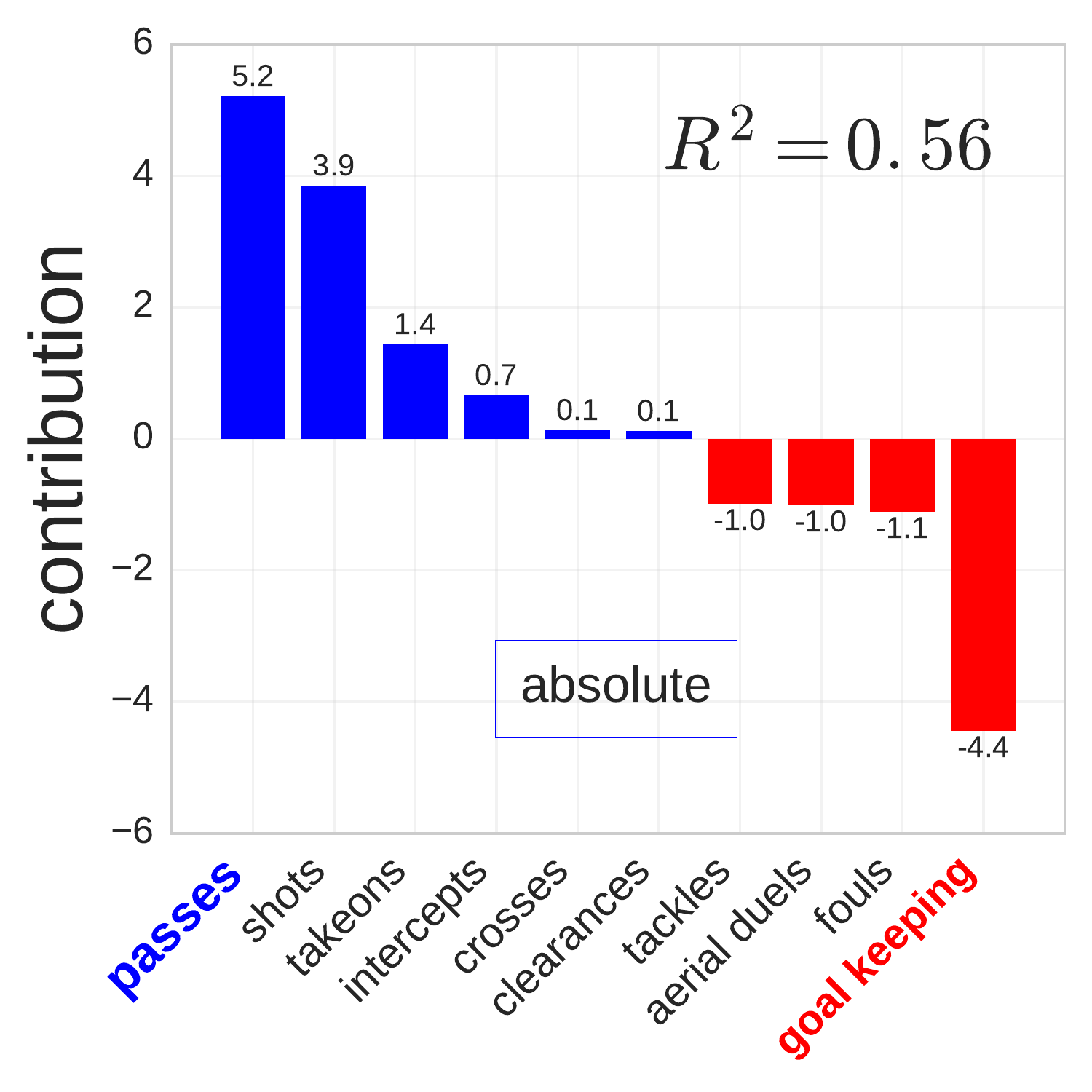}}
\subfigure[]{\includegraphics[scale=0.4]{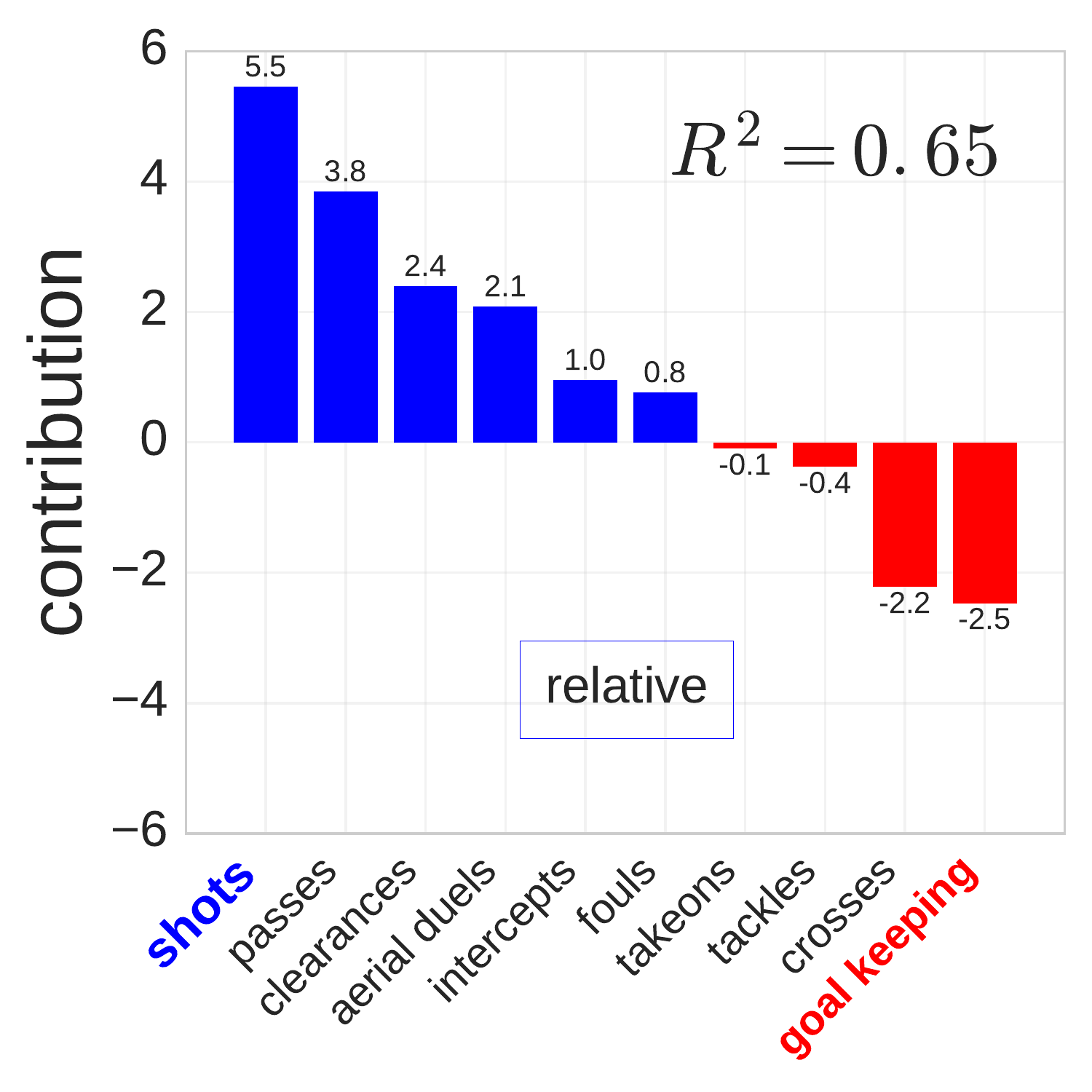}}
\caption{\textbf{Relative importance of performance features.} Coefficients produced by the OLS process for absolute performance features (a) and relative performance features (b).}
\label{fig:regr_importance}
\end{figure}

We move from a monodimensional view to a multidimensional view of performance and explore the predictive power of the combination of the single absolute performance features, by creating a model $M_{abs}$ via ordinary least squares (OLS).\footnote{We normalize every average performance feature $\overline{x}_i$ using the z-score normalization: $z = \frac{\overline{x}_i - \mu}{\sigma}$, where $\mu$ is the average and $\sigma$ the standard deviation of the feature's distribution. We implement OLS by using the \texttt{LinearRegression} object provided by the Python package \texttt{scikit-learn}.} We validate $M_{abs}$ using a 10-fold cross-validation scheme \cite{friedman2001elements}: the dataset is divided into 10 parts or folds, one fold is used as test set and the remaining folds as training set. Each sample in the dataset is tested once, using a model that is not fitted with that sample (see \ref{app:cross_validation}). Note that there is no need for considering the temporal ordering of games, since each team in a season is described by a single vector representing its typical performance.
The resulting cross-validated coefficient of determination is $R^2 = 0.56$, significantly better than the single features alone, indicating that multidimensional performance can explain more than half of the variance in the final score. By taking the normalized coefficients of the regression we can evaluate how strongly each feature influences a team's success. Figure \ref{fig:regr_importance}a shows the obtained coefficients indicating that the typical number of passes is the strongest driving force of success, more than the number of shots. In contrast, the number of goalkeeping actions negatively affect success, showing an absolute weight even higher than shots. 

We repeat the regression task by creating model $M_{rel}$ which uses the relative performance features. $M_{rel}$ produces better results than the absolute case as the resulting cross-validated coefficient of determination is $R^2=0.65$ (Table \ref{tab:pred_results}). This suggests that a team's success not only depends on how it performs typically, i.e., its playing style, but also on how it faces opponents, i.e., how it performs in relation to the opponent's playing style. Figure \ref{fig:regr_importance}b shows the coefficients of the typical relative features produced by model $M_{rel}$. In contrast with the absolute performance features, the difference in shots is the most important feature. In agreement with the absolute performance features, the difference in goalkeeping actions still contributes strongly and negatively (Figure \ref{fig:regr_importance}b). 
We also construct a regression model $M_{both}$ by using both the absolute and the relative performance features and observe no significant improvement with respect to $M_{rel}$. 
In Figure \ref{fig:regr_importance} we observe that some features like intercepts and crosses have a weight close to zero, i.e., they are almost irrelevant to the prediction. In fact, we repeat the regression experiments without those features and obtain similar results. 

We also train a logit model to classify each team, based on its performance features, in one of two classes of success: top teams promoted to a continental competition (\texttt{top}) and all the other teams (\texttt{bottom}).\footnote{Being promoted to a continental competition (i.e., Champions League or Europa league) provides a significant economic gain from the UEFA body as well as gain from sponsors and TV rights.} 
The ability of the logit model to generalize to teams that were not known during the training step provides us with a measure of how performance can discriminate between the two levels of success. For the absolute performance features we train a logit $C_{abs}$ and observe that it is significantly better than a baseline classifier which always predicts the most frequent class (i.e., \texttt{bottom}, see \ref{app:classification}). Similarly to the regression case, a classification model $C_{rel}$ trained on the relative performance features produces better results than the absolute performance case (\ref{app:classification}). These results indicate that the classifier can successfully rely on technical features to discriminate between the two classes of success.

\begin{table}[htb]\centering
\begin{tabular}{| >{\columncolor[gray]{0.95}}l | c | c | c | c || c | c | c | c |}
\cline{2-9}
\multicolumn{1}{c|}{} & \multicolumn{4}{c||}{\textbf{regression (OLS)}} & \multicolumn{4}{c|}{\textbf{classification (logit)}}\\
 \cline{2-9}
\multicolumn{1}{c|}{} & \multicolumn{2}{c|}{$M_{abs}$} & \multicolumn{2}{c||}{$M_{rel}$} & \multicolumn{2}{c|}{$C_{abs}$} & \multicolumn{2}{c|}{$C_{rel}$}\\
\hline
          \bf features &   $r$ &     $R^2$ & $r$ & $R^2$ &  \small ACC & F1 & \small ACC & F1\\
\hline
        tackles & -0.1 & -0.02 & -0.47 & \color{blue}0.19 & 0.67 &  0.54 & 0.72 & \color{blue}0.69 \\
   aerial duals & -0.14 & -0.01 & 0.37 & \color{blue}0.10 & 0.67 &  0.54 & 0.70 & \color{blue}0.66\\
        crosses & 0.12 & -0.01 & 0.45 & \color{blue}0.18 & 0.67 &  0.54 & 0.72 & \color{blue}0.70\\
  intercepts & -0.2 & 0.01 & -0.41 & \color{blue}0.13 & 0.67 &  0.55 & \color{blue}0.73 & \color{blue}0.70 \\
          fouls & -0.3 & 0.01 & -0.31 & 0.07 & 0.65 &  \color{blue}0.60 & 0.68 & \color{blue}0.60\\
        dribbles & 0.37 & 0.08 & 0.47 & \color{blue}0.19 & 0.69 &  \color{blue}0.65 & \color{blue}0.75 & \color{blue}0.72\\
     clearances & -0.36 &  0.09 & -0.34 &  0.09 & 0.70 &  \color{blue}0.65 & 0.69 & \color{blue}0.62 \\
     \hline
   goal keeping & -0.48 & \color{blue}0.22 & -0.69 & \color{blue}0.44 & \color{blue}0.74 &  \color{blue}0.72 & \color{blue}0.80 & \color{blue}0.79 \\
 shots & 0.67 & \color{blue}0.36 & 0.76 & \color{blue}0.54 & \color{blue}0.79 &  \color{blue}0.78 & \color{blue}0.84 & \color{blue}0.84 \\
     passes & 0.70 & \color{blue}0.45 & 0.71 & \color{blue}0.47 & \color{blue}0.80 &  \color{blue}0.79 & \color{blue}0.80 & \color{blue}0.80\\
     \hline
     \hline
     ALL & - & \color{blue}0.56  & - & \color{blue}\textbf{0.65} & \color{blue}0.82 & \color{blue}0.81 & \color{blue}\bf 0.86 & \color{blue}\bf 0.86\\
     \hline

\end{tabular}
\caption{\textbf{Results of regression and classification experiments using absolute and relative performance.} Pearson correlation coefficient ($r$), coefficient of determination ($R^2$), classification accuracy (ACC) and F1-score (F1) resulting from regression and classification experiments. We highlight in blue the models resulting in significant $R^2$ ($\ge 0.1$) or in accuracies and F1-scores with a significant improvement ($>0.05$) with respect to a baseline classifier which always predicts the most frequent class (for which ACC = $0.67$ and F1$=0.54$).}
\label{tab:pred_results}
\end{table}

\begin{figure}
\subfigure[]{\includegraphics[scale=0.4]{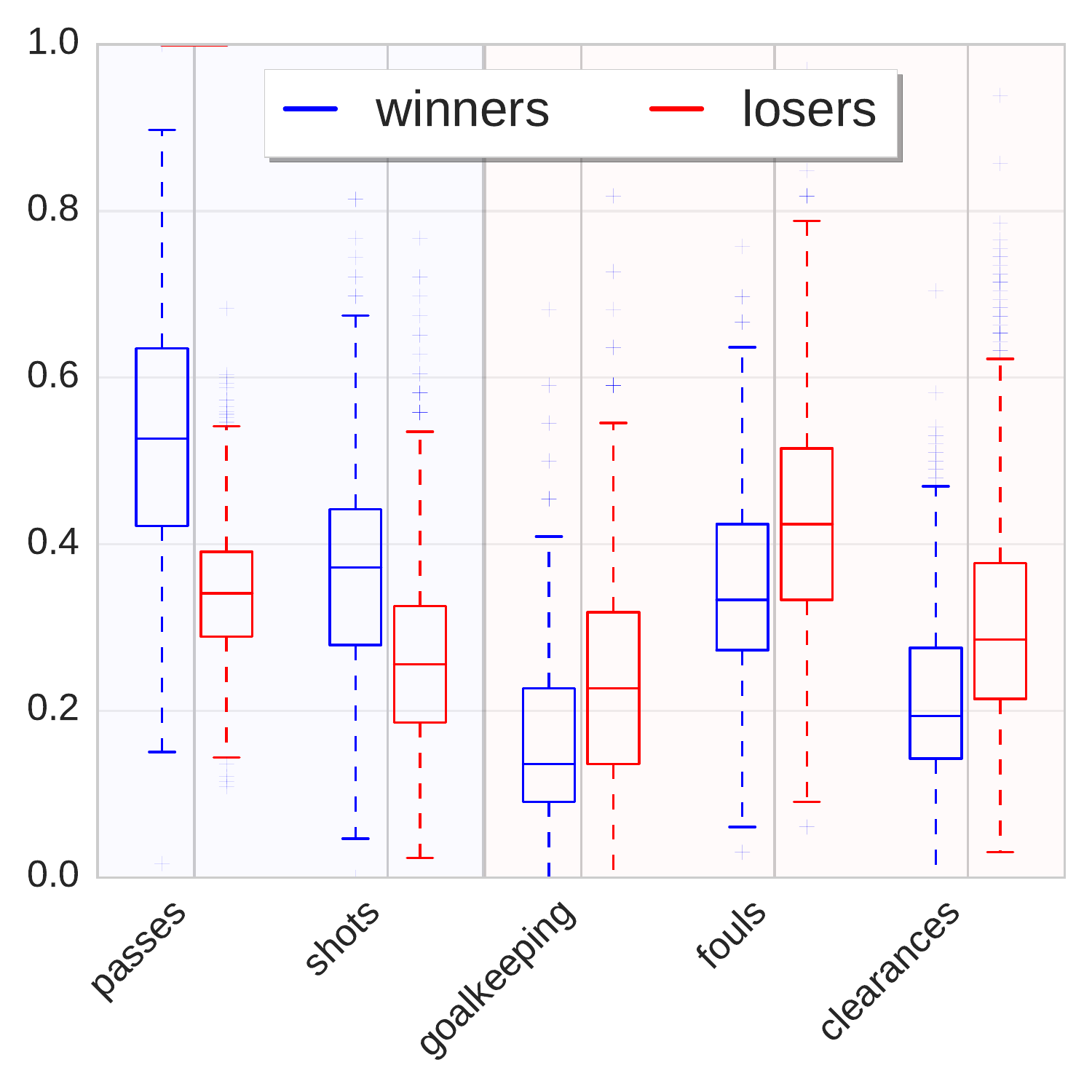}}
\subfigure[]{\includegraphics[scale=0.4]{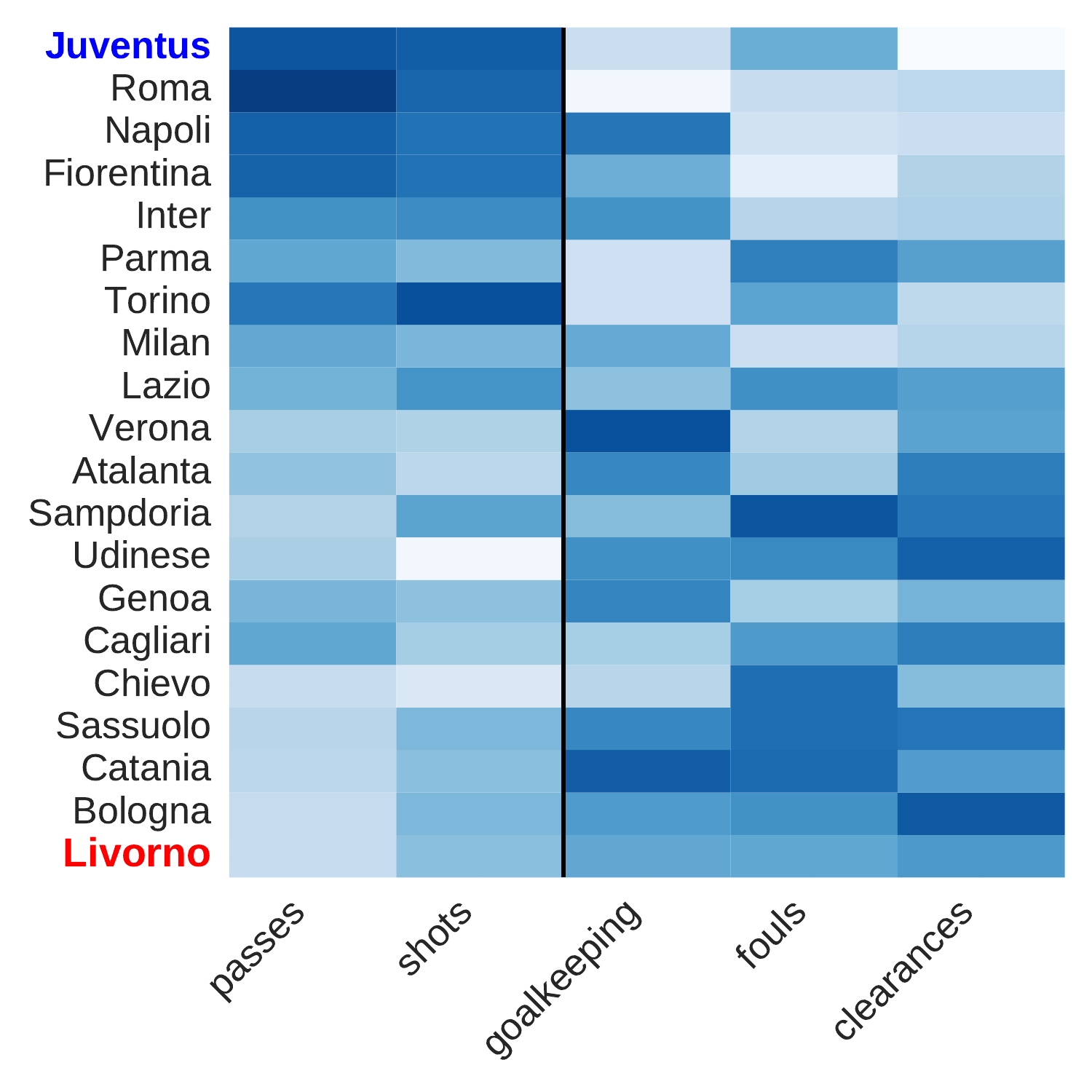}}
\caption{\textbf{Comparing the strongest and the weakest teams.} (a) Boxplots representing the distributions of five absolute performance features of winners (first in the final ranking) and losers (last in the final ranking). The values are normalized in the range [0, 1]. (b) Heatmap representing the normalized values (in the range [0, 1]) of five absolute performance features for all the teams in Serie A 2013/2014. Teams are sorted in decreasing order of points in the final ranking.}
\label{fig:example_perf}
\end{figure}

Figure \ref{fig:example_perf}a compares the distribution of five absolute performance features of winners (the first in the final ranking, in blue) and losers (the last in the final ranking, in red). The distributions significantly differ for all five features: in average, winners produce a higher number of passes and a higher number of shots than losers, suggesting that the most successful teams typically are dominant in both ball possession and attack opportunities. In contrast, losers suffer the dominance of the opponents and produce a higher number of goalkeeping interventions, fouls and defensive clearances. Figure \ref{fig:example_perf}b compares the typical values of the five performance features for all teams in Serie A in season 2013/2014, where the teams are sorted in decreasing order of points from the winner (FC Juventus) to the loser (AS Livorno). The high values of passes and shots are concentrated in the top of the ranking, while the high values of goalkeeping actions, fouls and clearances are concentrated in the bottom of the ranking. This result indicates that teams behave differently on the field according to their level of success.
 
\subsection{Validation against null models}
Regression and classification results show that a team's typical performance is related to its success. In order to test the significance of these relations we compare our findings with the results produced by two null models. 

\begin{figure}[h!]\centering
\subfigure[]{\includegraphics[scale=0.4]{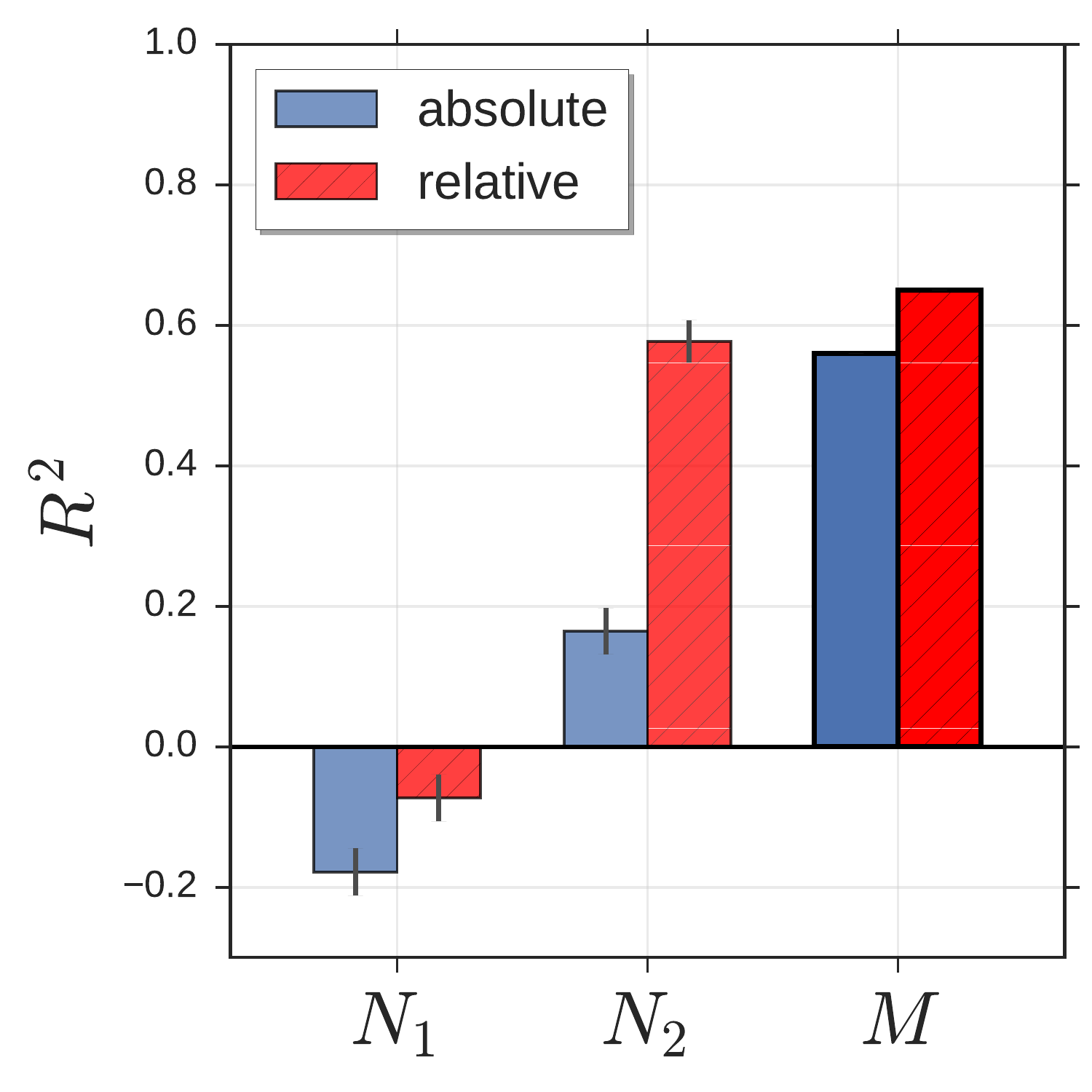}}
\subfigure[]{\includegraphics[scale=0.4]{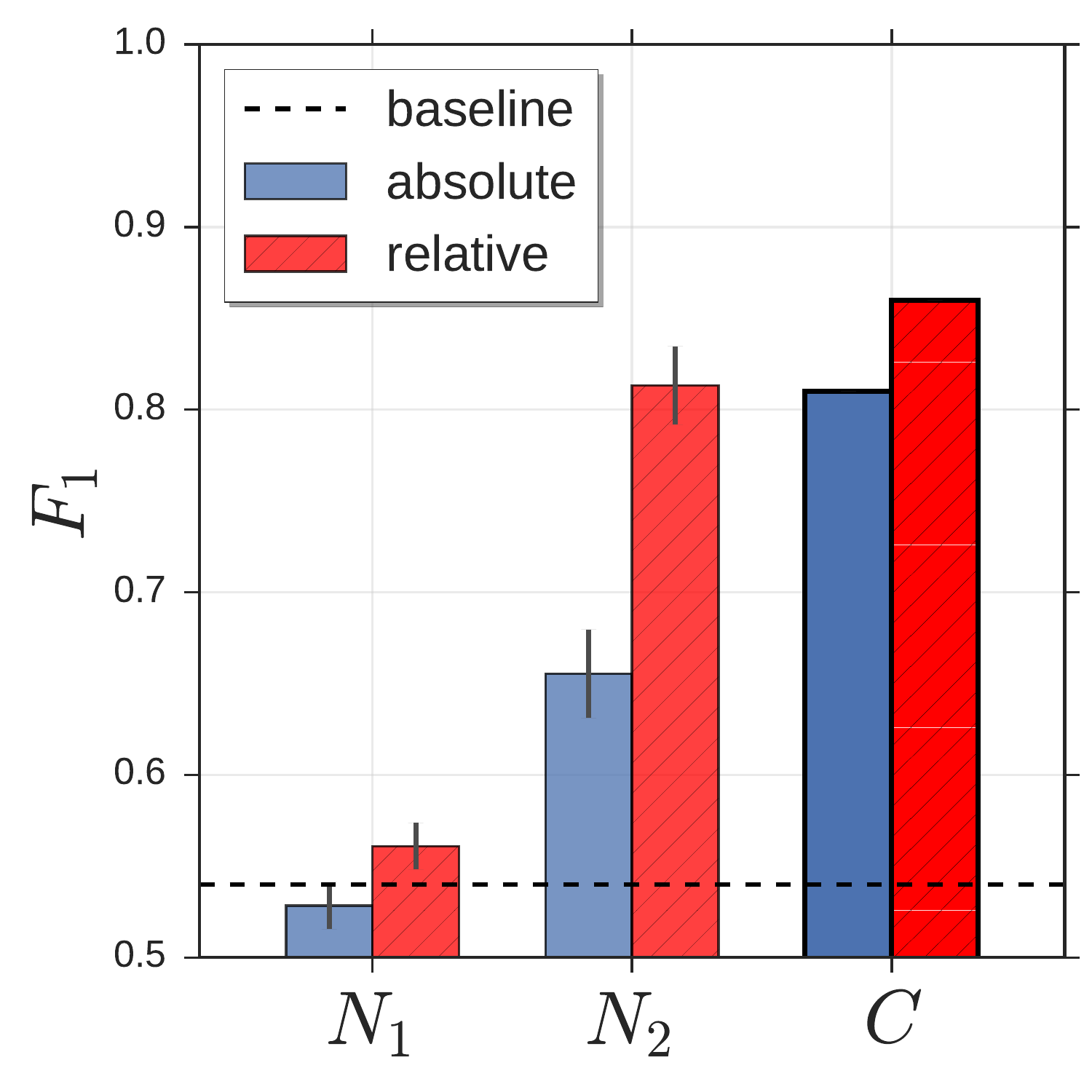}}
\subfigure[]{\includegraphics[scale=0.4]{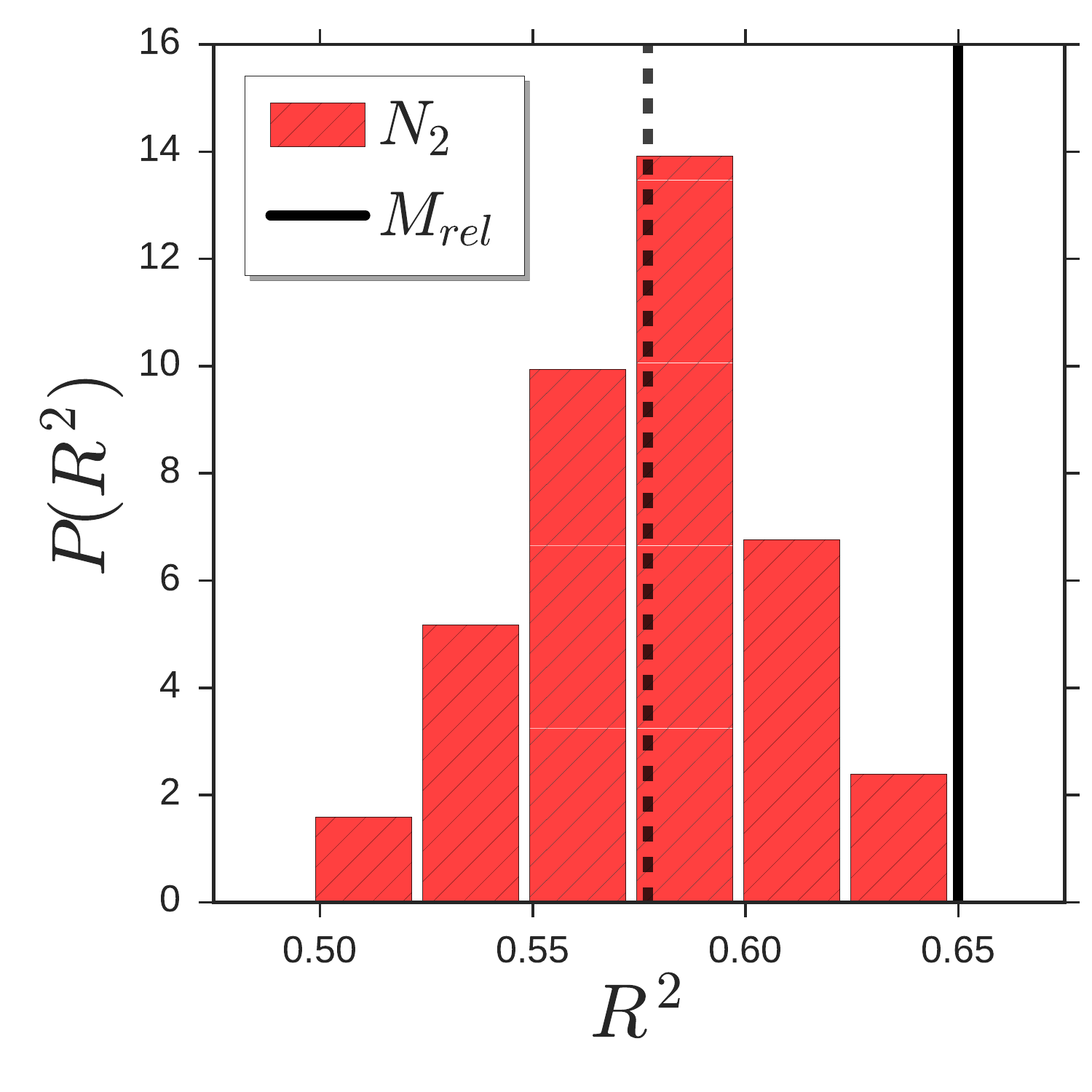}}
\subfigure[]{\includegraphics[scale=0.4]{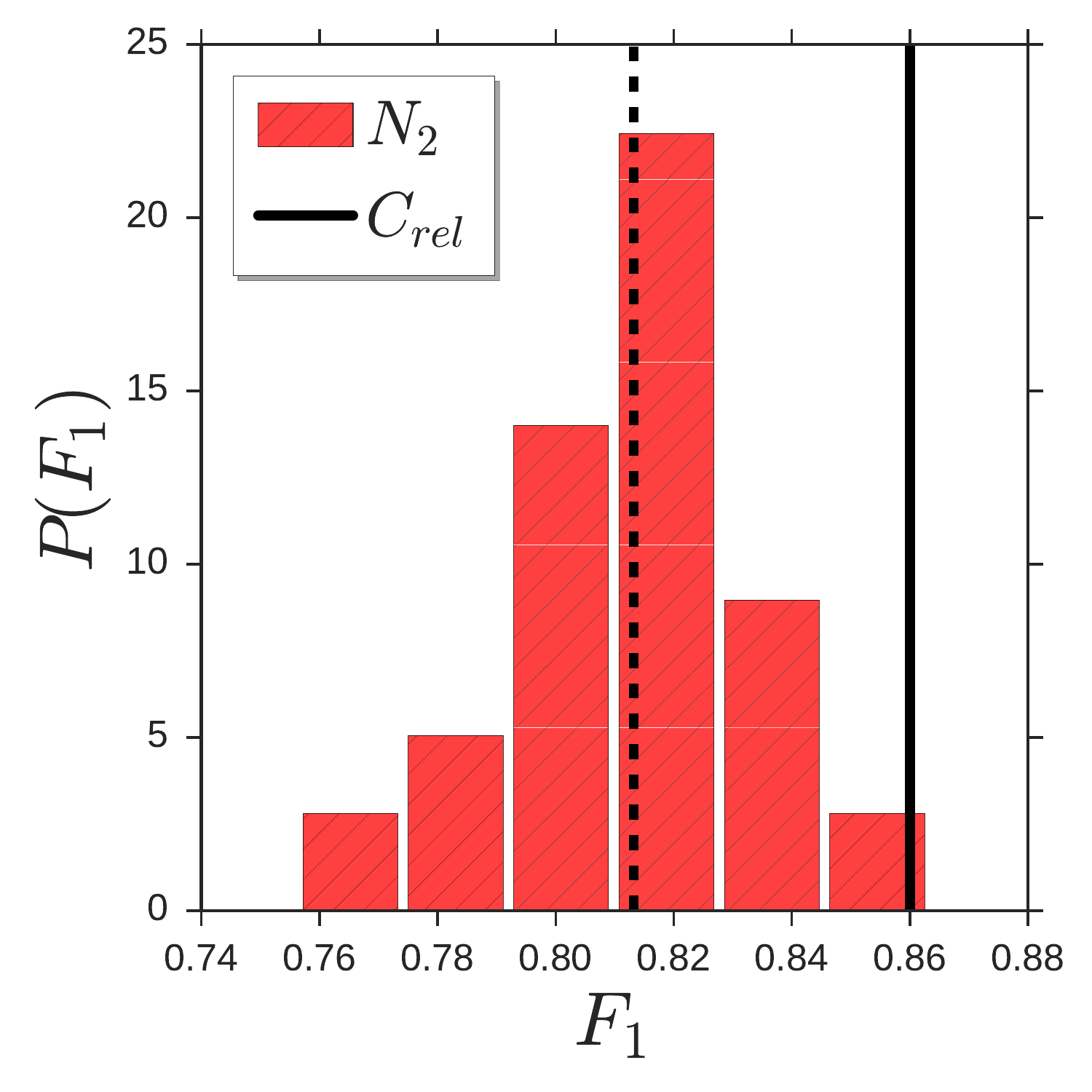}}
\caption{\textbf{Comparison of null models with empirical data.} (a) Comparison of the coefficient of determination ($R^2$) of null models $N_1$ and $N_2$ against empirical data ($M$), for absolute (blue) and relative (red) performance features. (b) Comparison of F-score of null models $N_1$ and $N_2$ against empirical data ($M$), for absolute (blue) and relative (red) performance features. (c-d) Distribution of coefficient of determination (c) and F-score (d) of null model $N_2$, compared to empirical data. The black dashed line is the average of the distribution of null model $N_2$.}
\label{fig:null_models}
\end{figure}

In null model $N_1$ we compute a team's typical performance by averaging the features over $n$ games chosen uniformly at random across all the games in the same competition and season, where $n$ is the number of games every team plays during the season. We then construct the regression and classification models to predict the final score and the level of success, respectively. Figure \ref{fig:null_models} compares the empirical $R^2$ and $F1$ of models with the average $R^2$ and the average $F1$ resulting from 1,000 runs of $N_1$. We find that, for both the absolute and the relative performance features, $N_1$ produces typical performances with no descriptive power with respect to success, resulting in $R_2{=}-0.17$ (a fit worse than a straight line) and $F1 = 0.54$ (as good as a classifier which always predicts the most frequent class). 

Null model $N_2$ computes a team's typical performance by choosing $n$ games at random preserving the outcome of every game, i.e., a team's actual game is replaced with another game with the same outcome (i.e., victories with victories, defeats with defeats, draws with draws). For the absolute performance features, $N_2$ produces typical performances resulting in $R^2 = 0.18$, better than $N_1$ but still far from the empirical data (Figure \ref{fig:null_models}a). For the relative performance features, $N_2$ achieves better predictive results producing $R^2=0.57$. This surprising result suggests that a team's typical relative performance is strongly related to the outcome of its games during the season and hence to its probability of winning a game. Hence we expect to observe a significant relation between relative performance and game outcome.

\begin{table}\centering
\begin{tabular}{|m{6cm} |m{6cm}|}
\multicolumn{2}{c}{\large Summary of results - Section \ref{sec:tournament_success}}\\
\hline
\centering \textbf{To what extent is performance descriptive of success?} & A regression model based on typical performance can explain up to 65\% of the variance in the final score. \\
\hline
\centering \textbf{To what extent is performance descriptive of a team's class of success?} & A machine learning classifier can accurately discriminate between top teams and bottom teams on the basis of the technical features.\\
\hline
\centering \textbf{What are the most important features for a team's success?} & The number of passes, shots and goalkeeping actions are the strongest predictor of a team's success.\\
\hline
\centering \textbf{What is the significance of the observed correlations?} & We validated the correlations against two different null models, which allows us to reject the hypothesis that our results occurred by chance \\
\hline
\end{tabular}
\end{table}

\section{Multidimensional performance and game outcome}
\label{sec:game_success}

Here we address the problem of detecting a team's victory, draw or defeat given its relative performance in a game. 
Previous works in the literature investigate the predictability of soccer results by modeling soccer tournaments as a stochastic process. For example, Heuer et al. \cite{heuer2009fitness} rely on the concept of team fitness to show that a soccer game can be described as two independent Poissonian processes. In this paper, we consider goal difference as a first evidence of success and investigate the technical aspects of a game which most determine it. We then create a game outcome predictor which approximates the relation between technical performance and goal difference to investigate the predictability of soccer games and perform a simulation of an entire season of six European leagues.

Given a team $A$ and a game $g$, we describe the problem of detecting a game outcome as a machine learning classification problem on the outcomes $\{1, 0, 2\}$, where 1 indicates a victory by $A$ (a positive goal difference), 0 indicates a draw (null goal difference) and 2 indicates $A$'s defeat (negative goal difference). 
We build a training set where each example is described by a label $r \in \{ 1, 0, 2\}$ indicating the team's outcome and a feature vector describing the team's relative performance in relation to the opponent. We try three alternatives to relate the performance of a team $A$ to the opponent $B$ in a game: \emph{(i)} we use $m=20$ performance features, $10$ features for $A$ and $10$ for the opponent $B$; \emph{(ii)} we use $10$ features where every feature $x_i = x_i(A) - x_i(B)$ is the difference between the $i$-th features of $A$ and the same feature of opponent $B$; \emph{(iii)} we use $10$ features where every feature $x_i = x_i(A) / (x_i(A) + x_i(B))$ is the ratio of the $i$-th features of $A$ over the sum of the feature of $A$ and opponent $B$. The three alternatives produce similar results, hence we use alternative $(ii)$ to limit the feature space and speed up the computations.

\begin{figure}\centering
\subfigure[]{\includegraphics[scale=0.4]{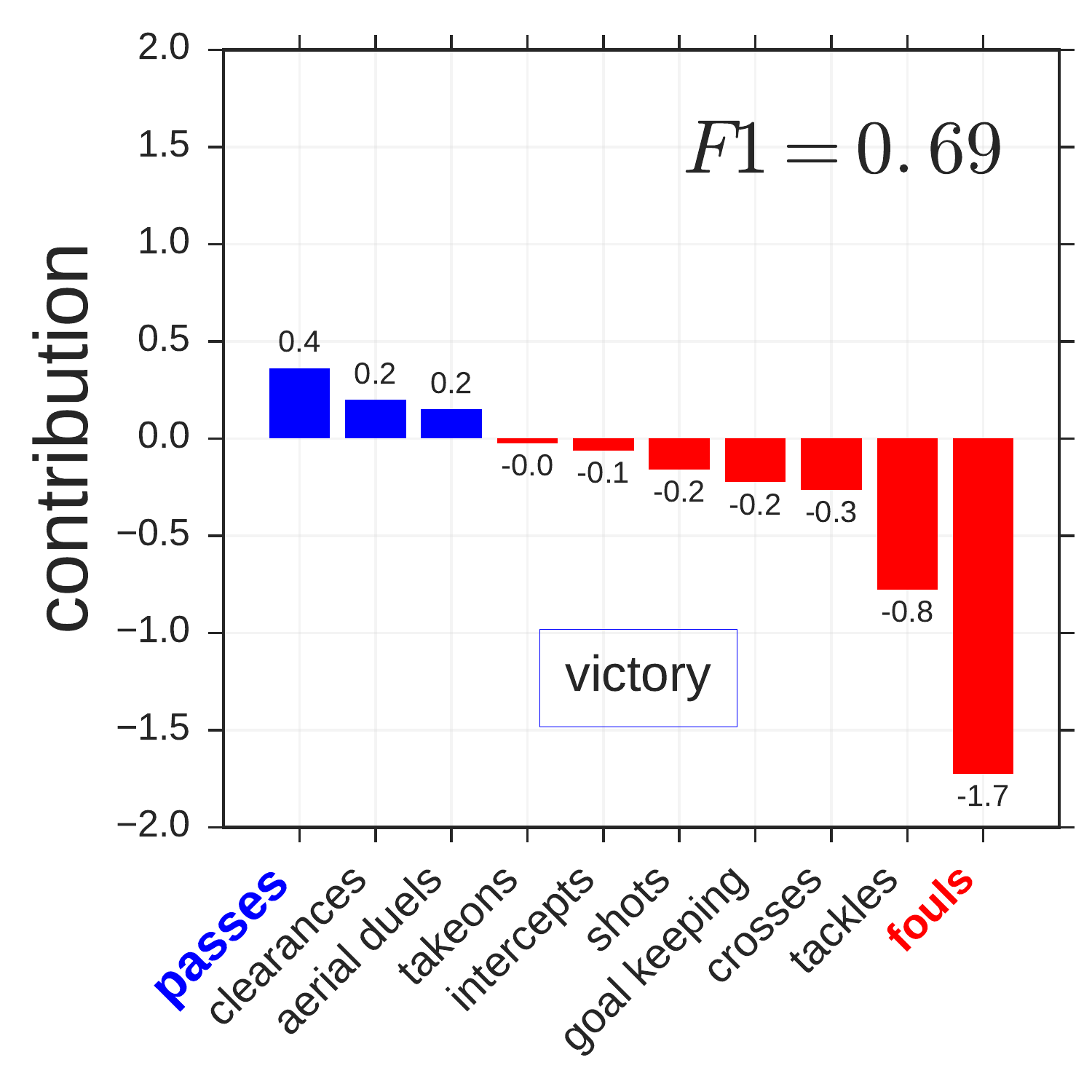}}
\subfigure[]{\includegraphics[scale=0.425]{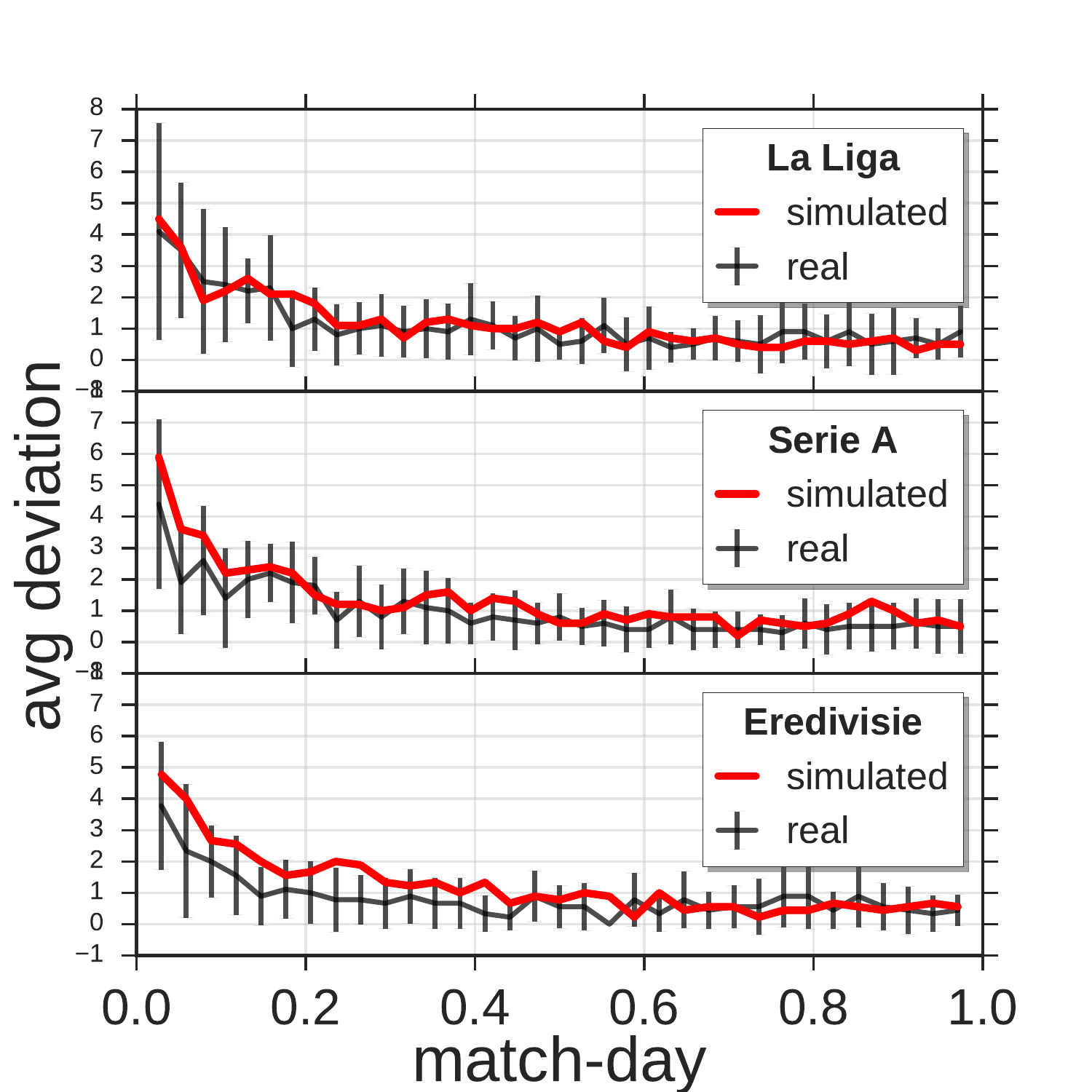}}
\caption{(a) \textbf{Relative importance of relative features in game outcome classification.} Coefficients produced by the logit $G_{rel}$ on the victory class. (b) \textbf{Average deviation of ranking positions over time, for three leagues.} At the beginning of the season, the deviation is high and it decreases as the season goes by. The deviation stabilizes at a typical value () from the half of the season. We normalize the match-day in the range [0, 1] to make leagues with different number of match-days (e.g., La Liga and Eredivisie) comparable.}
\label{fig:game_pred_importance}
\end{figure}

We perform the classification task by using a logit model $G_{rel}$ and evaluate it in terms of accuracy, $F1$, precision and recall (see Table \ref{tab:game_prediction}). We compare the classifier with a baseline which assumes no influence of the performance features on the outcome, i.e., it chooses a team's outcome at random according to the distribution of game outcomes in the training set. $G_{rel}$ has accuracy ACC${=}0.56$, meaning that it detects the outcome of a team in 56\% of the cases, significantly better than the baseline (ACC=$0.34$). Interestingly, while $G_{rel}$ can accurately predict a team's victory or defeat, it is difficult to detect draws (see Table \ref{tab:game_prediction}). The recall for the draw class is close to zero, indicating that just a few draws are detected by the logit. We also try a Random Forest classifier \cite{prinzie2007random}, which produces similar results in detecting draws. This result suggests that, while victory and defeat reflect a team's performance, draws cannot be fully explained, presumably because they are due to random or unpredictable episodes. We quantify the importance of every relative performance feature by taking the normalized coefficients produced by $G_{rel}$. We observe that producing more passes than the opponent is the major driver of a team's victory, as well as producing less fouls, tackles and goalkeeping actions during a game (Figure \ref{fig:game_pred_importance}a). In contrast, producing more fouls and less passes than the opponent can be symptom of a defeat (Figure \ref{fig:game_pred_importance}a).

\begin{table}[htb]\centering
\begin{tabular}{| >{\columncolor[gray]{0.95}}c | c | c || c | c | c | c | c | c |}
\cline{2-9}
 \multicolumn{1}{c|}{} & \multicolumn{2}{c||}{\textbf{overall}} & \multicolumn{2}{c|}{\textbf{victory} (1)} & \multicolumn{2}{c|}{\textbf{draw} (0)} & \multicolumn{2}{c|}{\textbf{defeat} (2)}\\
\hline
\textbf{models} & \textbf{acc} & \textbf{F} & \textbf{prec} & \textbf{recall} & \textbf{prec} & \textbf{recall} & \textbf{prec} & \textbf{recall}\\
\hline
logit & 0.60 & 0.52 & 0.60 & 0.81 & 0.00 & 0.00 & 0.60 & 0.81 \\
RF & 0.55 & 0.47 & 0.55 & 0.72 & 0.29 & 0.01 & 0.54 & 0.74\\
null model & 0.34 &  0.34 & 0.38 & 0.38 & 0.25 & 0.25 & 0.37 & 0.37\\
\hline
\hline
\texttt{logit gain} & \color{blue}0.26 & \color{blue}0.18 & \color{blue}0.22 & \color{blue}0.43 & \color{red}-0.25 & \color{red}-0.25 & \color{blue}0.27 & \color{blue}0.44\\
\hline
\texttt{RF gain} & \color{blue}0.21 & \color{blue}0.13 & \color{blue}0.17 & \color{blue}0.34 & \color{blue}0.04 & \color{red}-0.24 & \color{blue}0.17 & \color{blue}0.37\\
\hline
\end{tabular}
\caption{\textbf{Classification performance for game success prediction}. Accuracy, precision and recall of the classifiers on the static long-term scenario where the first half of the season is used as training set, while the second half of the season is the test set. Here the values are averaged across all the seven soccer leagues, since we do not find significant differences. RF = Random Forest.}
\label{tab:game_prediction}
\end{table}

\begin{table}\centering
\begin{tabular}{|m{6cm} |m{6cm}|}
\multicolumn{2}{c}{\large Summary of results - Section \ref{sec:game_success}}\\
\hline
\centering \textbf{What is a game outcome predictor?} & It is a machine learning classifier which assigns a game outcome (0, 1 or 2) to a team's relative performance.\\
\hline
\centering \textbf{To what extent is the game outcome predictor accurate?} & It accurately discriminates between victories and defeats, while it is difficult to detect draws.\\
\hline
\centering \textbf{What are the key feature for a team's game outcome?} & Producing more passes than the opponent is the major driver of a team's victory, as well as producing less fouls, tackles and goalkeeping actions. Producing more fouls and less passes than the opponent can lead to a defeat.
 \\
\hline
\end{tabular}
\end{table}

\section{Tournament simulation and PC ranking}
\label{sec:simulation}
Model $G_{rel}$ synthesizes the relation between technical performance and game outcome. We use it to simulate the evolution of an entire season of the six national tournaments and construct a ranking of all teams in a competition -- the PC ranking.
First, we train $G_{rel}$ on the games in seasons 2013/2014 and 2014/2015. Then, we use it to simulate, match-day by match-day, the game outcomes in season 2015/2016. Depending on the game outcome predicted by $G_{rel}$, every team gains three points for a victory, one point for a draw and no points for a defeat. Finally we compute the PC ranking at the end of a competition as the teams' cumulative number of points. We compare a competition's PC ranking with its actual ranking by two metrics: \emph{(i)} the correlation between the teams' points in the two rankings; \emph{(ii)} the accuracy of defining the groups of success (top five, bottom five, all the rest), computed as the ratio of teams in the PC ranking which resulted to be in their actual group of success. To test whether each competition has its peculiar relation between performance and game outcome, we repeat the simulation for each competition separately by using as training set the games of that competition only. Table \ref{tab:simulation_results} shows the similarity between PC rankings and actual rankings. 

\begin{table}[htb]\centering
\def\arraystretch{1.5}%
\begin{tabular}{l | r | r || r | r |}
\cline{2-5}
& \multicolumn{2}{c||}{\bf all} & \multicolumn{2}{c|}{\bf single} \\ \hline
\multicolumn{1}{|l|}{\bf{tournament}} & \bf corr   & \bf acc  & \bf corr   & \bf acc \\ \hline
\multicolumn{1}{|l|}{Bundesliga}          & 0.64                 & 0.62     & 0.69 & 0.55 \\\hline
\multicolumn{1}{|l|}{Premier League}      & 0.56                 & 0.50  & 0.44                 & 0.55            \\ \hline
\multicolumn{1}{|l|}{La Liga}                & \cellcolor{blue!20}\textbf{0.71}        & 0.55   & 0.68                 & 0.50          \\ \hline
\multicolumn{1}{|l|}{Eredivisie}          & 0.68                 & 0.40    & 0.62                 & 0.55          \\ \hline
\multicolumn{1}{|l|}{Serie A}             & 0.62                 & 0.60     & \cellcolor{blue!20}\textbf{0.74}        & \cellcolor{blue!20}\textbf{0.70}         \\ \hline
\multicolumn{1}{|l|}{Ligue 1}             & 0.56                 & \cellcolor{blue!20}\textbf{0.80}  & 0.43                 & 0.55   \\ \hline
\end{tabular}
\caption{\textbf{Similarity between PC rankings and actual rankings.} Correlation and group accuracy between PC rankings and actual rankings where the game outcome predictor model is trained on all the competitions indiscriminately (all) and only on the seasons of a specific competition (single).}
\label{tab:simulation_results}
\end{table}

We find a significant similarity in both the scenarios and for both the metrics. In particular, the scenario where we train the game outcome predictor on games from all the competitions produces better results than the single competition scenario, with the exception of Serie A. We obtain the best PC ranking for La Liga, with a Pearson correlation $r=0.71$. Serie A shows the best group accuracy: 80\% of teams in the PC ranking result in their actual group of success. 

We also compare the PC rankings with the teams' Elo ratings at the end of the season. Elo is a standard algorithm to rank players and teams according to their recent outcomes \cite{lasek2013predictive,Hvattum2010460}. A team's Elo rating at match-day $i$ is determined by the team's Elo rating after match-day $i - 1$ and the game outcome 
in match-day $i$. It is worth highlighting a crucial difference between Elo and our simulation: while Elo ranks the teams on the only basis of their recent results, our simulation exploits the relationship between the performance of the two teams and the game's outcome, as inferred from the machine learning model trained on the previous season. Our simulation, hence, aims at demonstrating that the technical features can explain part of a team's success during a season, while Elo provides a summary of the team's recent success. We compute the Elo ratings in the following way. Given a season $j$, we first compute the Elo ratings for all teams in season $j-1$ \cite{lasek2013predictive}. This provides us with the teams' initial ratings at the beginning of season $j$.\footnote{In season $j-1$, the initial rating is 1500 for all the teams \cite{lasek2013predictive,Hvattum2010460}.} We then compute the Elo ratings for season $j$, updating them match-day by match-day according to the actual game outcomes \cite{lasek2013predictive,Hvattum2010460}. At the end, the Elo ratings provide a measure of the teams' strength according to their game outcomes during the season.

Tables \ref{tab:spanish_league_ranking} compare the actual ranking, the PC ranking and the Elo ranking for La Liga 2015/2016. We observe a good agreement between the PC ranking and the actual ranking especially for the teams at the top: the winner of the tournament (FC Barcelona) is correctly identified by the simulation with an error of just 3 points, and three on four teams qualified to the Champions League are predicted in the exact group (FC Barcelona, Real Madrid and Atl\'etico Madrid). Although the Elo ratings can correctly predict the winner, it is less accurate on identifying the position of the other teams qualified to the Champions League. For some teams, the PC ranking error is high, meaning that the simulation overestimates or underestimates the success they achieve in the competition (Figure \ref{fig:errors}). For example in the PC ranking Levante got 16 points more than it actually achieved, while Sevilla got 16 points less. 
Figure \ref{fig:errors} shows, for every team in La Liga, the difference between the points in the actual ranking and the points in the PC ranking. The simulation tend to overestimate the number of points of the top teams and to underestimate the number of points of the bottom teams. The differences between the PC ranking and the actual ranking can be related to the fact that draws are to some extent unpredictable and hence often misclassified. Even after averaging over a whole season, differences between the two rankings can still remain since draws are around 26\% of the games. Elo ratings have a higher correlation with the points in the actual ranking ($r = 0.80$) w.r.t.\ the PC ranking ($r = 0.71$). However, the Elo ratings' coefficient of variation $cv_{elo}$ is much lower that the coefficient of variation in the actual ranking ($cv_{\tiny \mbox{Elo}}= 1\%$, $cv_{\tiny \mbox{actual}} = 33\%$), while the coefficient of variation $cv_{PC}$ in the PC ranking is comparable to the actual ranking one (i.e., $cv_{\mbox{\tiny PC}}=31\%$).\footnote{The coefficient of variation of a distribution is the ratio between the standard deviation and the mean of the distribution.} 

We also investigate the impact of random and systematic effects in determining a team's strength, by investigating the variation of rankings as the season goes by \cite{heuer2009fitness}. For each league we compute the average variation of the teams' positions in the ranking, match-day by match-day: at a given match-day $i$ we compute for each team its absolute difference $d_i(A) = |p_i(A) - p_{i - 1}(A)|$, where $p_i(A)$ indicates team $A$'s position in the ranking after match-day $i$ and $p_{i - 1}(A)$ the position after match-day $i-1$. Figure \ref{fig:game_pred_importance}b shows how this absolute difference changes as the season goes by, for the actual rankings and the PC rankings. The two curves show a similar behavior: while at beginning of the season we observe a high average variation in the rankings position, this variation decreases as the season goes by, stabilizing at around half of the season. A team's final position in the ranking emerges hence in the long run, both in the actual rankings and in the PC rankings. This suggests that, despite the substantial unpredictability of draws, unexpected results do not significantly affect the final rankings in a competition. 

\begin{table}[htb]\centering

\begin{tabular}{lr | lr | lr}

\multicolumn{2}{c|}{\bf actual ranking} &   \multicolumn{2}{c|}{\bf PC ranking} &   \multicolumn{2}{c}{\bf ELO ratings} \\
\hline
         \bf \color{blue}Barcelona &   \bf \color{blue}91 &    \bf \color{blue}Barcelona & \bf \color{blue}94 & \color{blue}\bf Barcelona & \bf \color{blue}1465.24\\
         
          \color{blue}Real Madrid &           \color{blue}90 & \color{blue}  Real Madrid &\color{blue} 77 & \color{blue} At\'etico Madrid & \color{blue}1462.75\\
          
          \color{blue}Atl\'etico Madrid&           \color{blue}88 &  \color{blue}   Atl\'etico Madrid &  \color{blue}  69 & \color{blue} Real Madrid & \color{blue}1462.12\\
          
        \color{blue}Villarreal &           \color{PineGreen}64 &           \color{PineGreen}Celta Vigo &          \color{blue}69 & Valencia & 1452.50 \\
        
           \color{PineGreen}Athletic Bilbao & \color{PineGreen}62 & Las Palmas &          69 
           & \color{PineGreen}Sevilla & \color{PineGreen}1450.12\\
           
            \color{PineGreen} Celta Vigo &        \color{PineGreen}   60 &          Deportivo &          65 
            & \color{PineGreen} Athletic Bilbao & \color{PineGreen}1435.25\\
            
          \color{PineGreen} Sevilla &      \color{PineGreen} 52 &             \color{blue} Villareal &          \color{blue} 61 
          &  \color{blue}Villareal &  \color{blue}1433.87\\
          
            Malaga &           48 &        Betis &        61 
            & Real Sociedad & 1429.25\\
            
          Real Sociedad &           48 &            Granada &          55  &
          \color{PineGreen}Celta Vigo & \color{PineGreen}1428.87\\
          
             Betis &           45 &          \color{PineGreen}Athletic Bilbao & \color{PineGreen} 51 
             & M\'alaga & 1426.50\\
             
          Valencia &           44 &           Real Sociedad &          49 
          & Espanyol & 1422.75  \\
          
        Las Palmas &           44 &           Valencia &          48 
        & Vallecano  & 1419.87\\
        
             Eibar &           43 &        \color{red}    Levante &      \color{red}    48 
             & Granada & 1417.00\\
             
          Espanyol &           43 &             Malaga &          47 
          & Deportivo & 1415.25\\
          
         Deportivo &           42 &           Espanyol &          47 
         & \color{red}Getafe & \color{red}1415.00\\
         
         Vallecano &           41 &              Eibar &          46 
         & \color{red}Levante & \color{red}1414.37\\
         
 Sporting Gijon &           39 &  Sporting Gijon &          42 &
Eibar & 1411.50\\
 
           Granada &           39 &         \color{PineGreen} Sevilla & \color{PineGreen}  36 
           & Las Palmas & 1399.62\\
           
         \color{red}   Getafe &        \color{red}   36 &      \color{red}       Getafe &     \color{red}     26 & Betis & 1398.75\\
         
         \color{red}  Levante &        \color{red}   32 &          Vallecano &          20 
         & \color{red}Sporting Gij\'on & \color{red}1395.62\\
\hline
 & & \multicolumn{2}{c|}{$r=0.71$} & \multicolumn{2}{c}{$r = 0.80$}\\
 \multicolumn{2}{c|}{$cv_{\tiny \mbox{actual}} = 33\%$} & \multicolumn{2}{c|}{$cv_{\tiny \mbox{PC}}=31\%$} & \multicolumn{2}{c}{$cv_{\tiny \mbox{Elo}} = 1\%$}\\
\end{tabular}
\caption{\textbf{Actual ranking, PC ranking, and Elo ratings of La Liga 2015/2016.} Teams in blue are qualified to Champions League, teams in green are qualified to Europa League, teams in red are relegated in the second division.}
\label{tab:spanish_league_ranking}
\end{table}

\begin{table}\centering
\begin{tabular}{|m{6cm} |m{6cm}|}
\multicolumn{2}{c}{\large Summary of results - Section \ref{sec:simulation}}\\
\hline
\centering \textbf{What is PC ranking and how it is computed?} & PC ranking is a team ranking emerging from our tournament simulation. We use the game outcome predictor to assign an outcome to a game, given the teams' performances. The cumulative points gained by the teams according to the synthetic outcomes compose the PC ranking.\\
\hline
\centering \textbf{How similar the PC rankings are to the actual rankings?} & For La Liga 2015/2016, we find a Pearson correlation $r=0.76$ between the PC rankings and the actual ranings, as well as similar and a ranking coefficients of variation (PC=$31\%$, actual=$33\%$).\\
\hline
\centering \textbf{How similar the Elo ratings are to the actual rankings?} & For La Liga 2015/2016, we find a Pearson correlation $r=0.88$ between the Elo ratings and the actual ranking. However, the ratings' coefficients of variation are very different (Elo=$1\%$, actual=$33\%$).
 \\
\hline
\end{tabular}
\end{table}

\section{Discussion}
\label{sec:discussion}

Our analysis reveal four main results. First, a team's typical performance is significantly related to its success, as we observe $R^2 = 0.65$ (Table \ref{tab:pred_results}). We confirm these results against two different null models, an observation that allows us to reject the hypothesis that our discovery occurred by chance (Figure \ref{fig:null_models}). This result is particularly valuable when considering that we use only 10 features to describe technical performance.
Five features matter the most to the observed relation: ball possession, shots, goalkeeping interventions, fouls committed and defensive clearances (Figure \ref{fig:regr_importance}). These features allow us to discriminate between the top teams and the bottom teams in a competition, as well as to characterize how the most successful teams behave on the field with respect to the least successful ones (Figure \ref{fig:example_perf}).

Second, we find that relative performance provides a better descriptive power than absolute performance (Table \ref{tab:pred_results}). Since two teams influence one another during a game, their behavior should be put in relation to each other. Ball possession seems to be a notable exception to this rule: absolute and relative passes show a similar correlation with success (Table \ref{tab:pred_results}). An interpretation of this result is that ball possession is strongly related to the amount of time a team controls the game. Since time is limited, producing a high number of passes automatically implies producing more passes than the opponent. 

Third remarkable result is that, while victories and defeats can be explained by technical performance, draws are difficult to detect (Table \ref{tab:game_prediction}). Previous works in the literature show that draws are difficult to predict and, as a consequence, odds for draws are similar for basically all soccer matches \cite{heuer2013perfekte,heuer2009fitness}. Our results link the observed unpredictability to technical performance and demonstrate that draws do not correspond to well-defined technical behaviors.

Finally, we find a surprising similarity between the competitions' actual rankings and the PC rankings produced by our simulations (Table \ref{tab:simulation_results} and Table \ref{tab:spanish_league_ranking}). This result indicates that a game outcome predictor, which properly synthesizes the relation between performance and game outcome, can be successfully used to explain a team's progress during a competition. Our simulation, and the related PC ranking, can be useful tools to understand to what extent a team's success reflects its performance on the field. Part of the divergence between the actual and the simulated points observed for some teams (e.g., Levante in Table \ref{tab:spanish_league_ranking} and Figure \ref{fig:errors}) can be due to the fact that draws are largely unpredictable, and even after averaging over a whole season significant unpredictable effects remain. Nevertheless, the observed divergence can be also due to the limited number of features considered, or by the existence of contextual or psychological factors that are not measured by existing technologies. As our results suggest, while the success of some teams (e.g., FC Barcelona in Table \ref{tab:spanish_league_ranking}) can be accurately explained by their performance, the success of other teams (e.g., Levante in Table \ref{tab:spanish_league_ranking}) relies on factors that are either not captured by soccer logs or non measurable by existing technologies.

\section{Conclusion}
\label{sec:conclusion}
In this work, we analyzed 6,396 games and 10 million events in the top six European leagues and take a further step towards the understanding of the complex relation between performance and success in soccer, a sports where these two quantities can be individually measured. As we show, a model based on this relation can be used to define alternative ranking systems for soccer teams and point out to what extent a team's success reflects its technical behavior on the field.

Our work can be extended in several directions. First, soccer logs can be combined with tracking data, which describe the spatio-temporal trajectories of players during a game \cite{DBLP:journals/corr/GudmundssonH16,stein2017how}. Tracking data describe aspects of a game that are not captured by soccer logs, adding significant information. Second interesting direction is the developing of a mathematical model to generate synthetic data describing the performance of two opposing teams. This model should embeds the relation between performance and success and reproduce game patterns in an accurate way. Finally, while in this paper we focus on team performance, it would be interesting to study the problem focusing on players and detect which technical features influence the success of a player in a game or competition. In the meanwhile, experiences like ours may contribute to shape the discussion on how to predict success from Big Data, such as soccer logs, that are massively available nowadays. If we learn how to exploit such a resource, we have the potential of creating systems in support of coaches, managers and practitioners which can rely on data-driven simulations to boost a team's performance and predict its future success.

\appendix

\section{Soccer leagues format}
\label{app:round_robin}
During a league each of the $n$ participating clubs plays against each of the other clubs twice, once at home and once away, for a total of $n(n-1)$ games. The season is split into two halves. In the first half of the season each club plays once against each league opponent, for a total of $n(n - 1)/2$ games. In the second half of the season the clubs play in exactly the same order that did in the first half of the season, the only difference being that home and away situations are switched. The games are organized in $2n - 2$ match-days. All the games in match-day $i$ are played before the games in match-day $i + 1$, even tough some games can be anticipated or postponed to facilitate players and clubs participating in Continental or Intercontinental competitions. In all the leagues, clubs are awarded three points for a victory, one point for a draw, and no points for a defeat. At the end of the season, the winner of the league is the club with the most points, and other clubs can be qualified to continental competitions according to specific rules defined by the UEFA.

\section{Descriptive statistics of the soccer dataset}
\label{app:distr}
\begin{figure}\centering
\includegraphics[scale=0.40]{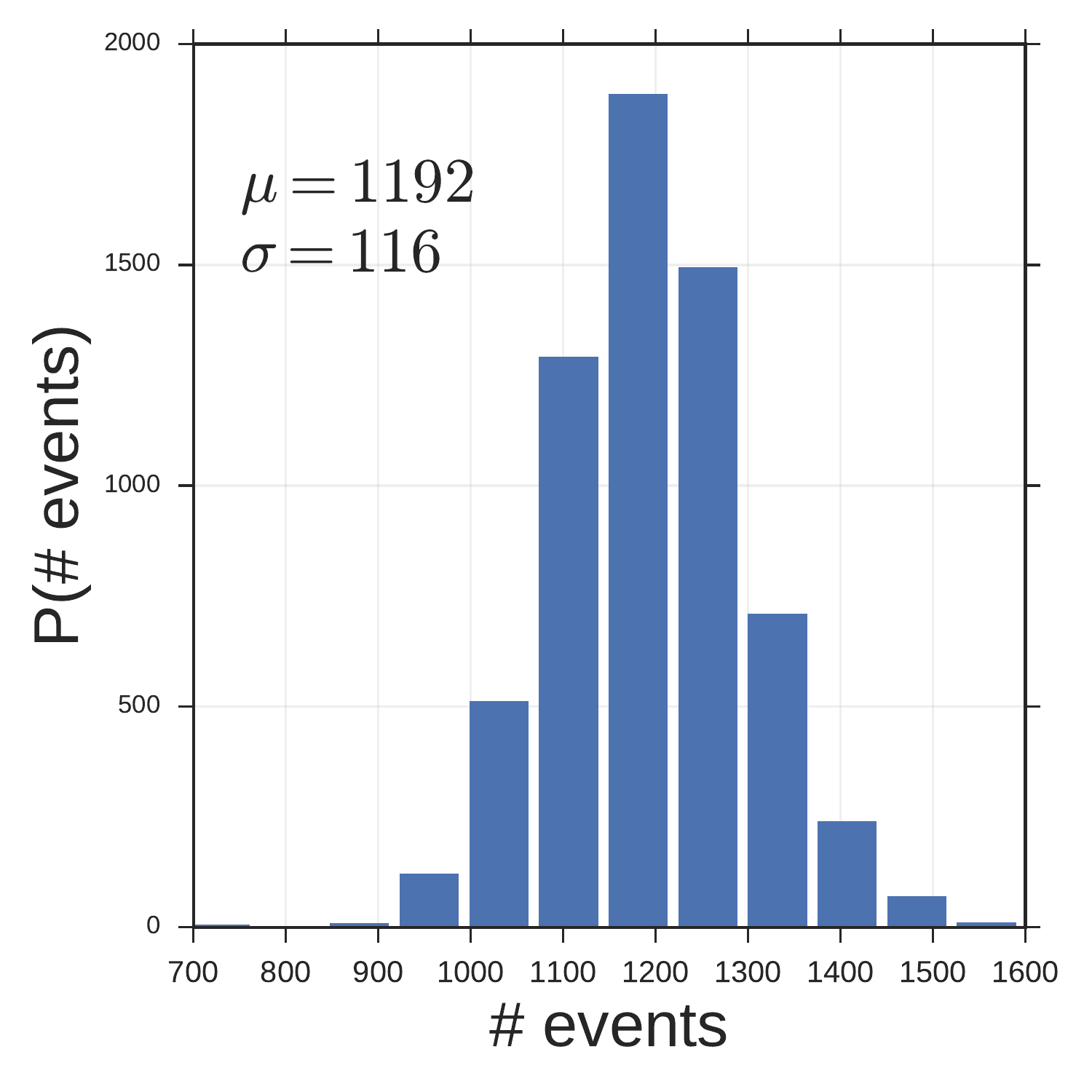}
\includegraphics[scale=0.40]{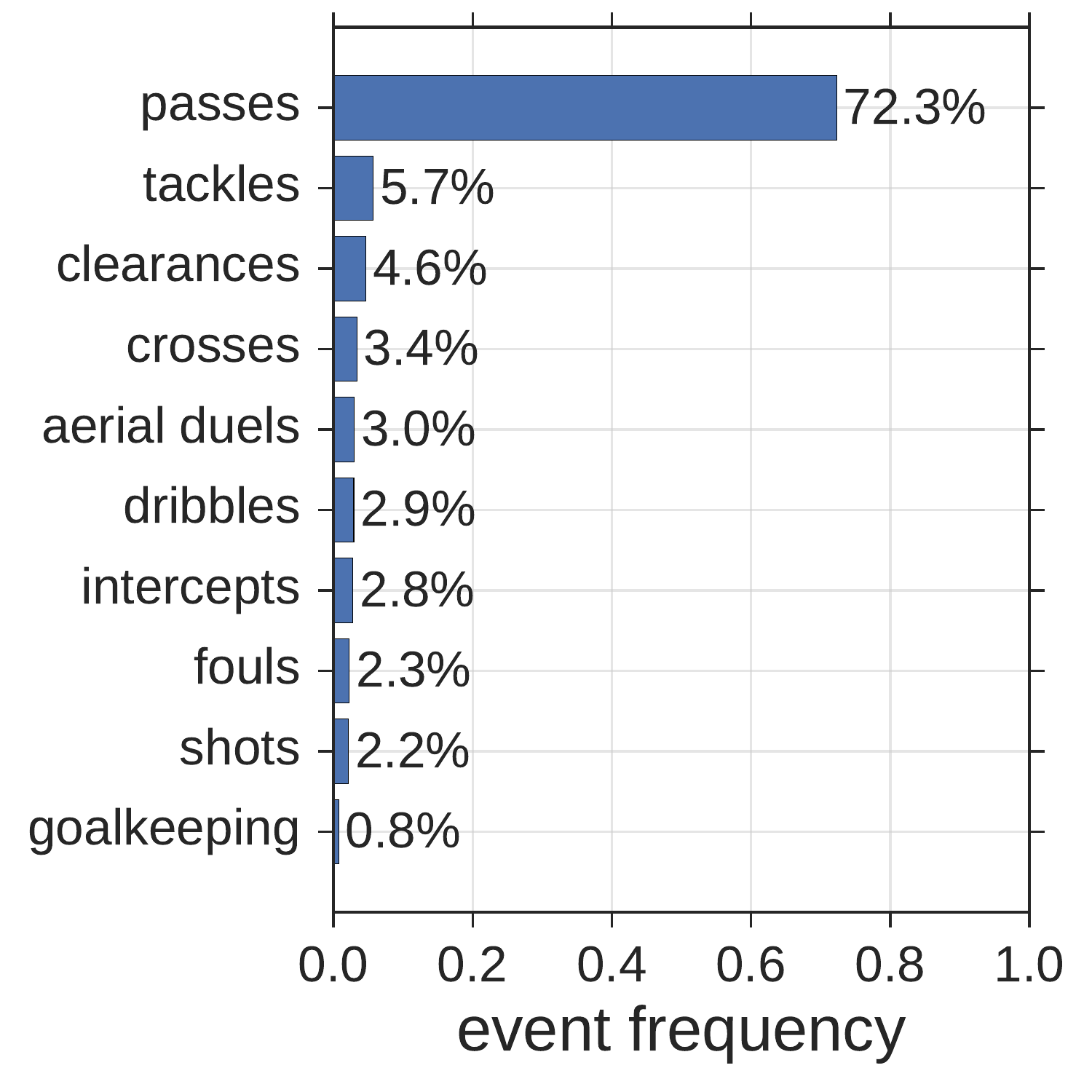}
\caption{\textbf{The distribution of the number of events in soccer games.} (a) Distribution of the number of events per game across all the games in the dataset. We observe a normal distribution with average $\mu= 1,192$ and standard deviation $\sigma \pm 116$. (b) Frequency of the different types of events occurring during the games in the dataset. We observe that passes are the most frequent events (72.3\% of the events).}
\label{fig:distr_events}
\end{figure}

Figure \ref{fig:distr_events}a shows the distribution of the number of events per game. The distribution is well fitted by a normal distribution, describing that a typical soccer game produces around $\mu = 1,192$ events in average with a standard deviation of $\sigma \pm 116$, hence denoting a coefficient of variation $\sigma/\mu \approx 10\%$. Figure \ref{fig:distr_events}b shows the proportion of event types across all the games in our dataset. We observe that passes are the most frequent events,
 accounting for more than 70\% of the events. Tackles are the second most frequent events, followed by clearances, crosses, aerial duels, dribbles, intercepts, fouls, shots and goalkeeping actions. Goals, which are the most important events for the outcome of a game, are the rarest ones accounting for just the 0.2\% of the events.
 
\section{Quality and spatial performance features}
\label{app:quality_features}

The ten metrics used in the paper are directly extracted as pure counts from the ten event types available in soccer logs: passes, crosses, shots, tackles, dribbles, clearances, goalkeeping, actions, fouls, intercepts and aerial duels. Considering just pure counts can ignore other information that is important for our purpose. For this reason we also extract a set of features describing a team's ``quality'' on each soccer event and a set of features describing a team's spatial and temporal ``dominance'' during a game. In particular, we consider the following features:
\begin{itemize}
\item \texttt{pass precision}, the ratio between a team's completed passes and the team's total number of passes during a game. The higher the pass precision, the better is the team's pass efficacy;
\item \texttt{dribble precision}, the ratio between the dribbles completed by the team and the dribbles attempted by the team;
\item \texttt{tackle precision}, the ratio between the tackles completed by the team and the tackles attempted by the team;
\item \texttt{corners}, the number of crosses that are performed as corner kicks;
\item \texttt{cross precision}, the ratio between the number of crosses that reach the destination player and total number of crosses;
\item \texttt{attack/defense}, the ratio between a team's number of shots and the number of its goalkeeping actions (a proxy of attack vs defense attitude).
\end{itemize}
\begin{itemize}
\item \texttt{average position}: the team's average position on the field during the game;
\item \texttt{average attack position}: the team's average position when performing attack events (crosses, shots and dribbles);
\item \texttt{average defensive position}: the team's average position when performing defensive events (tackles, clearances, intercepts, fouls);
\item \texttt{total play actions}: the total number of play actions created by the team;
\item \texttt{total play action duration}: the total duration of the play actions created by the team;
\item \texttt{average speed}: the average speed of the play actions created by the team, measured as the ratio between the distance covered and the action duration;
\item \texttt{average acceleration}: the average acceleration of the team's play actions, where every acceleration is computed as $$\frac{(100 - \mbox{distance from goal})^2}{\mbox{passes}}.$$
\end{itemize}

\begin{figure}\centering
%first row
\subfigure[]{\includegraphics[scale=0.27]{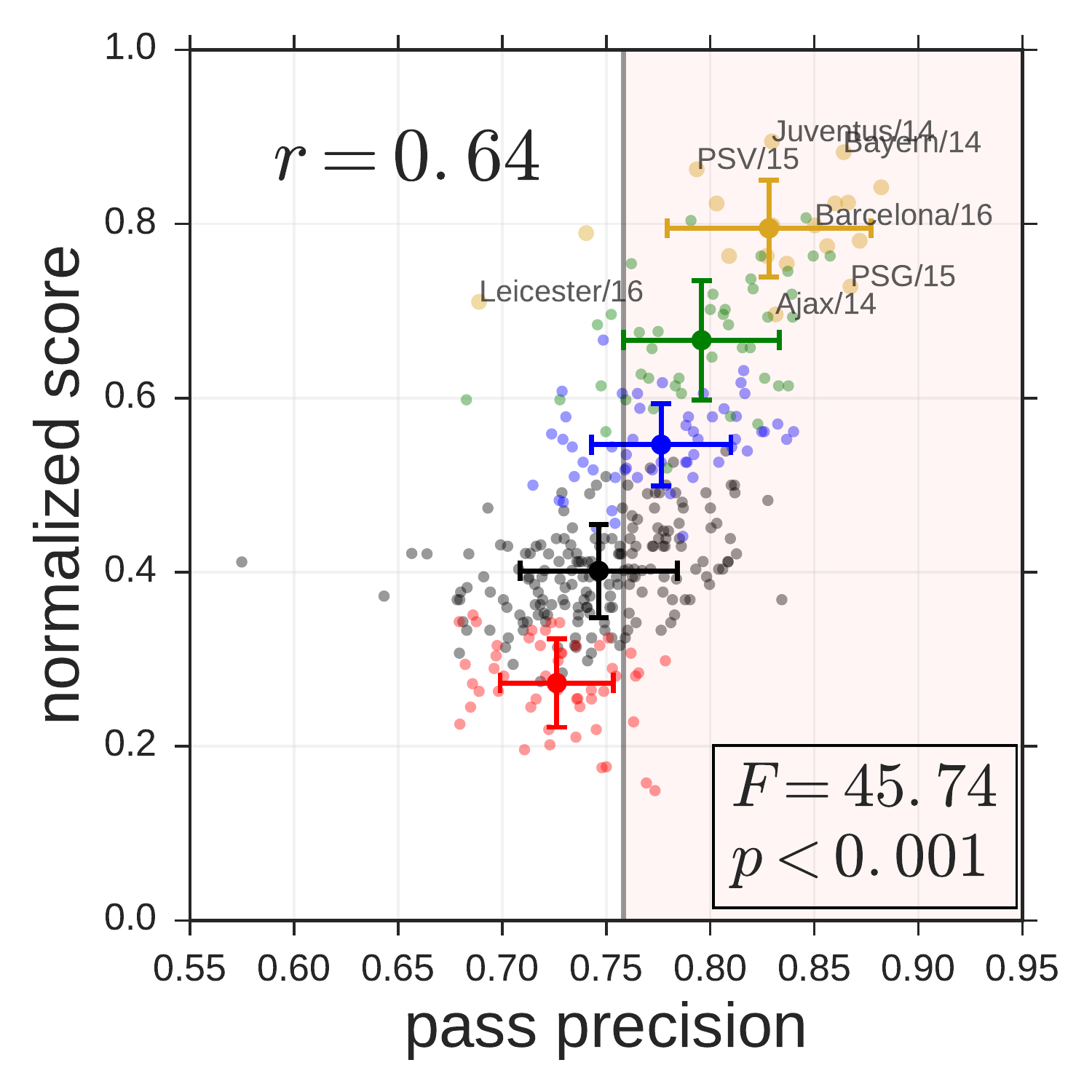}}
\subfigure[]{\includegraphics[scale=0.27]{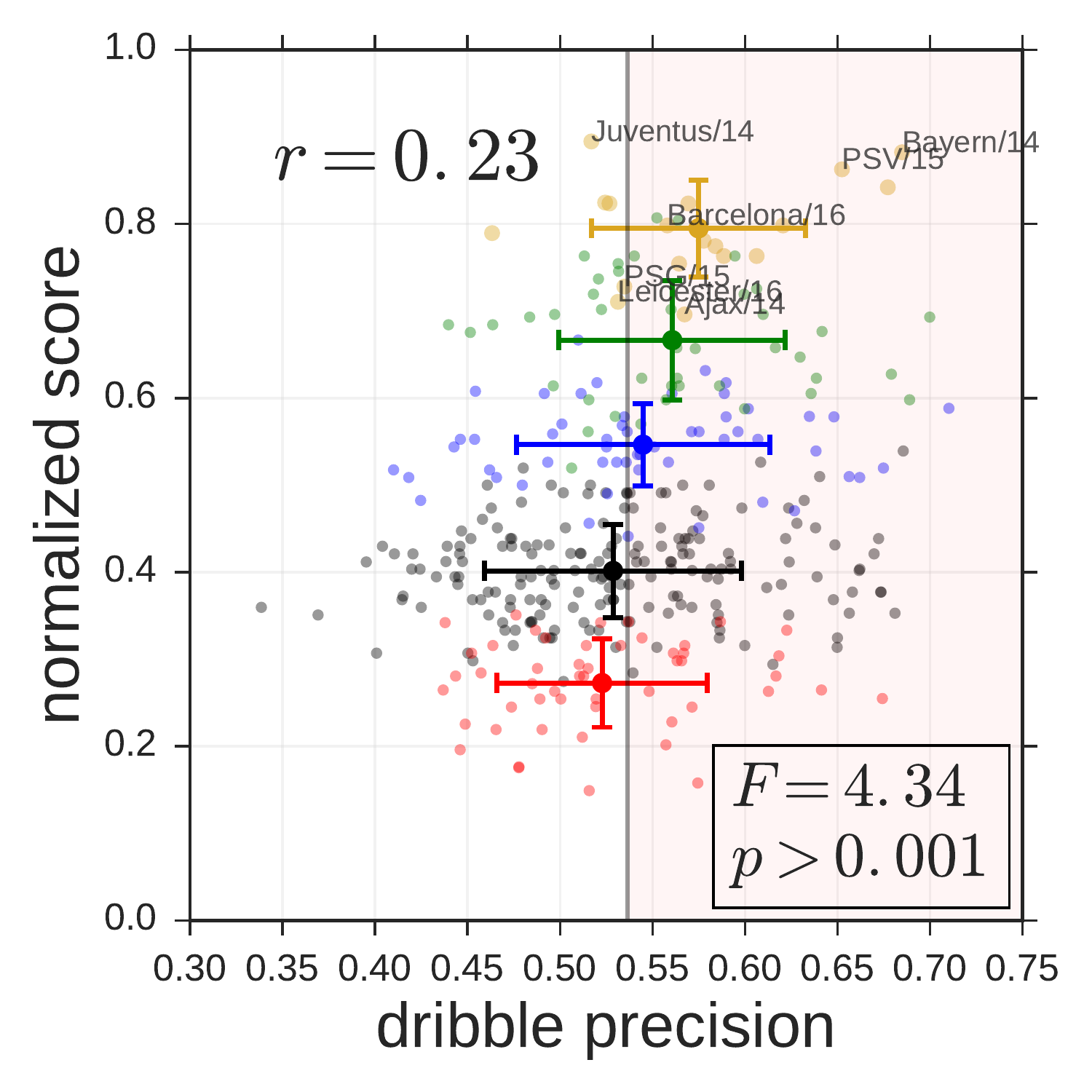}}
\subfigure[]{\includegraphics[scale=0.27]{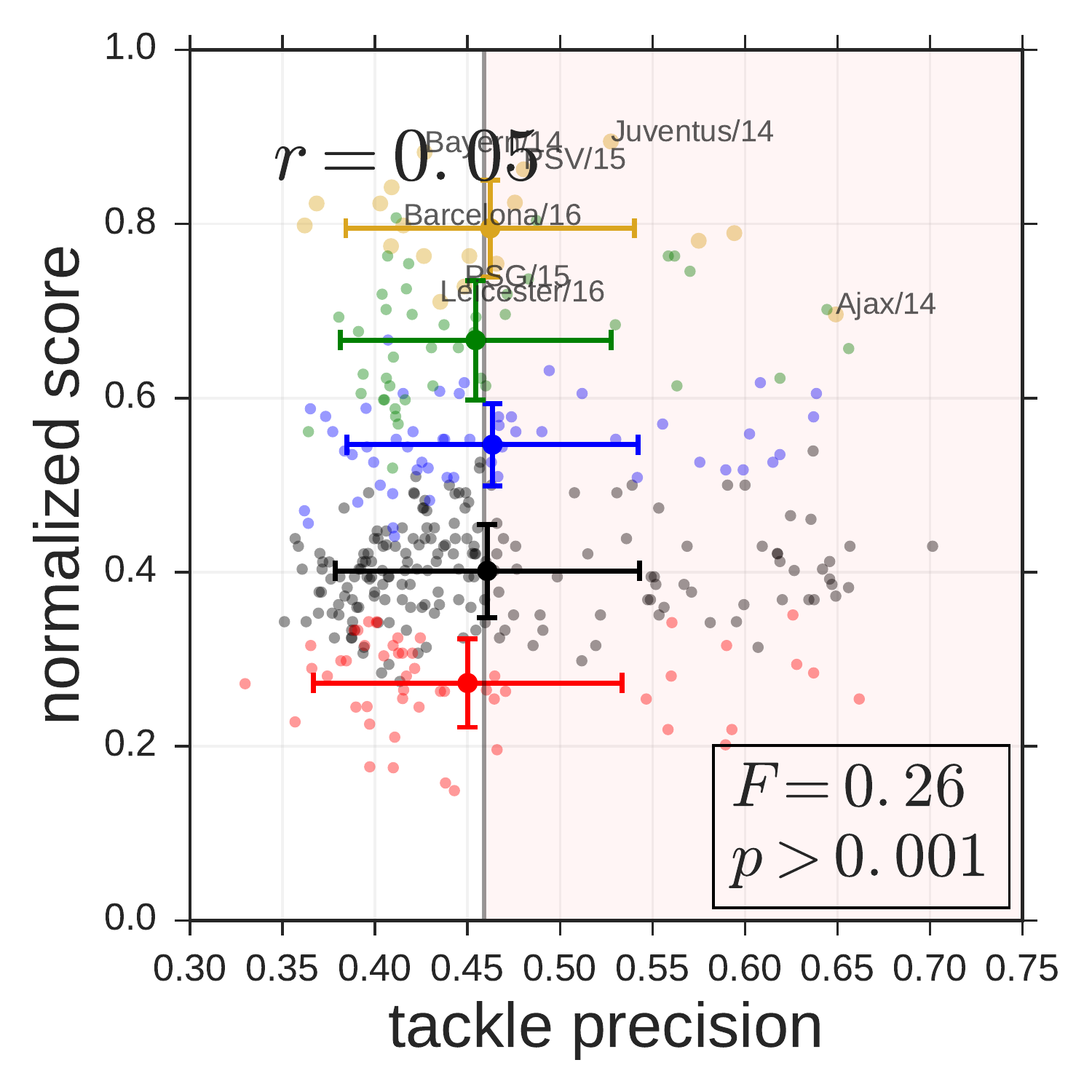}}

%second row
\subfigure[]{\includegraphics[scale=0.27]{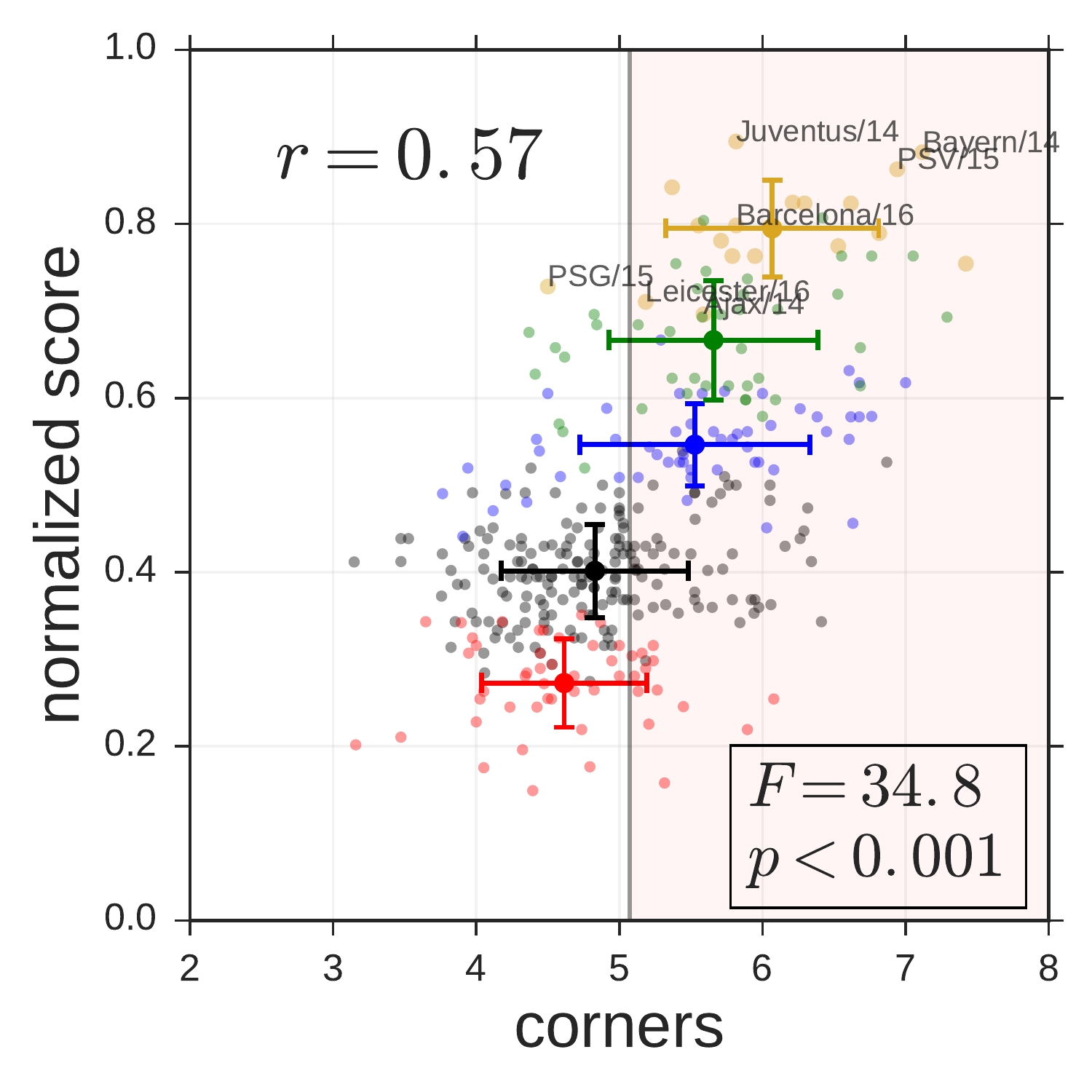}}
\subfigure[]{\includegraphics[scale=0.27]{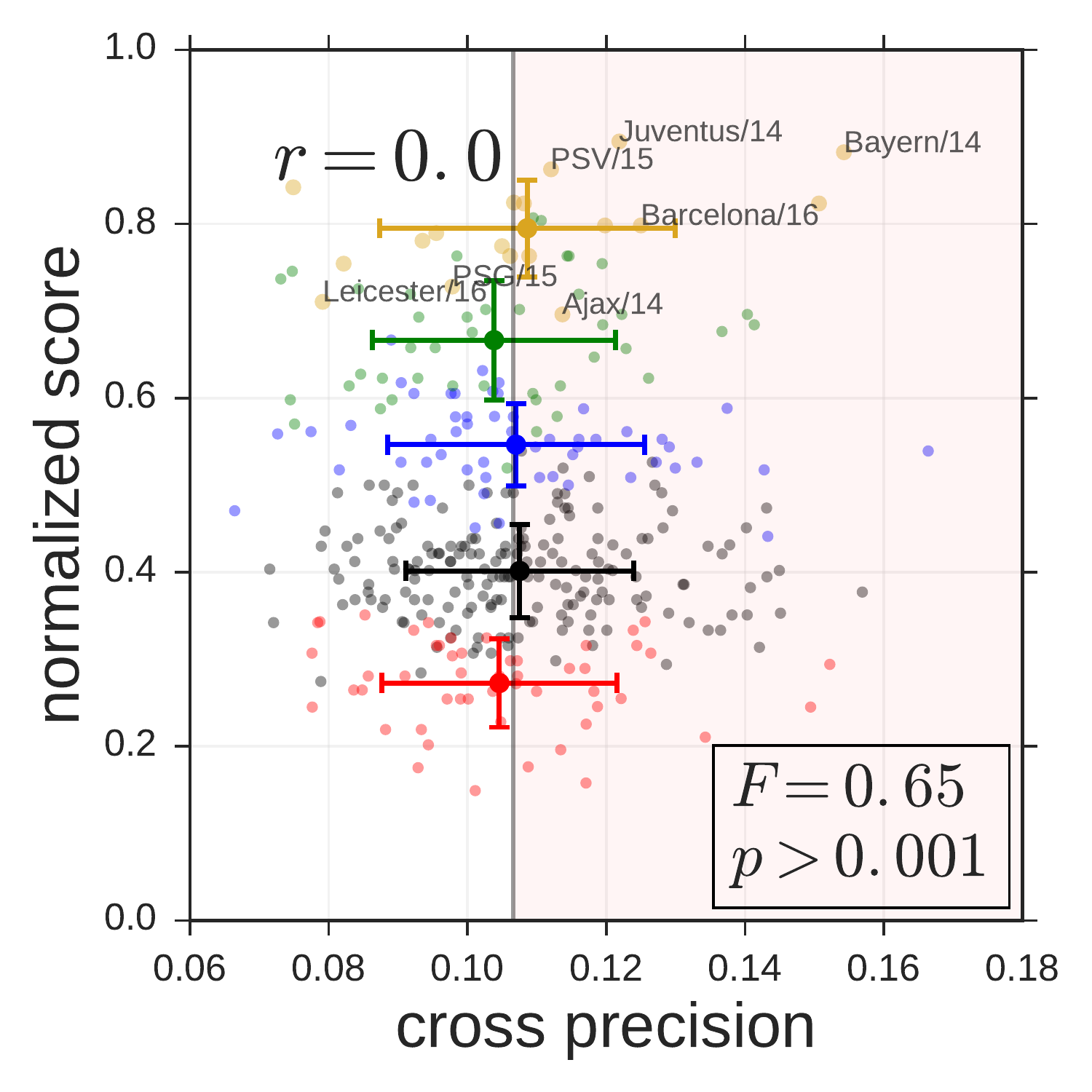}}
\subfigure[]{\includegraphics[scale=0.27]{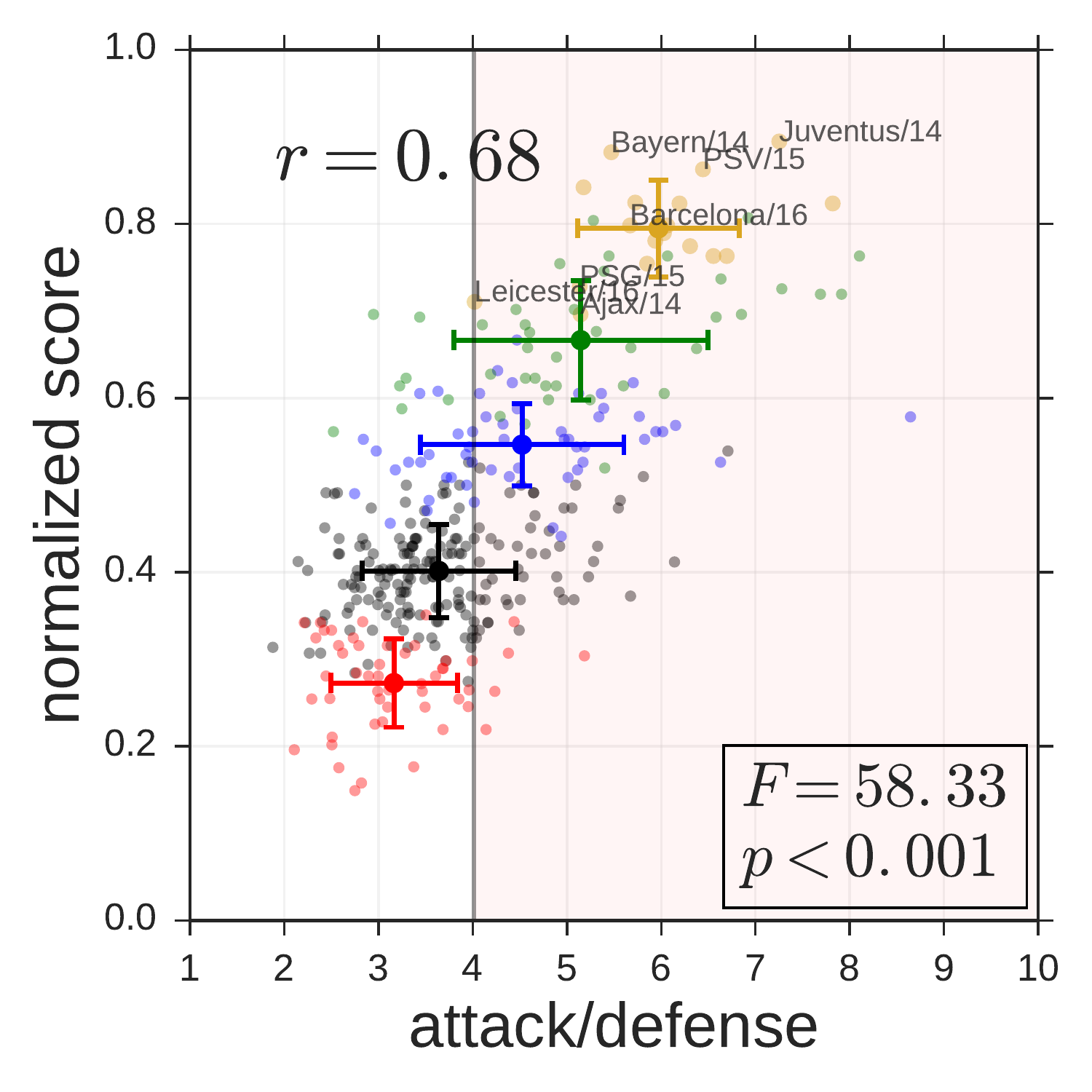}}

\caption{\textbf{Correlation between quality features and success}. We split the teams in groups according to the final ranking: teams relegated in second division (red), teams in the middle (black), teams in Europa League (blue), teams in Champions League (green) and winners (golden). In the box we indicate F-test statistic and p-value resulting from a one-way ANOVA to assess whether the values of features within the groups differ from each other.}
\label{fig:new_corrs}
\end{figure}

\begin{figure}\centering
%first row
\subfigure[]{\includegraphics[scale=0.27]{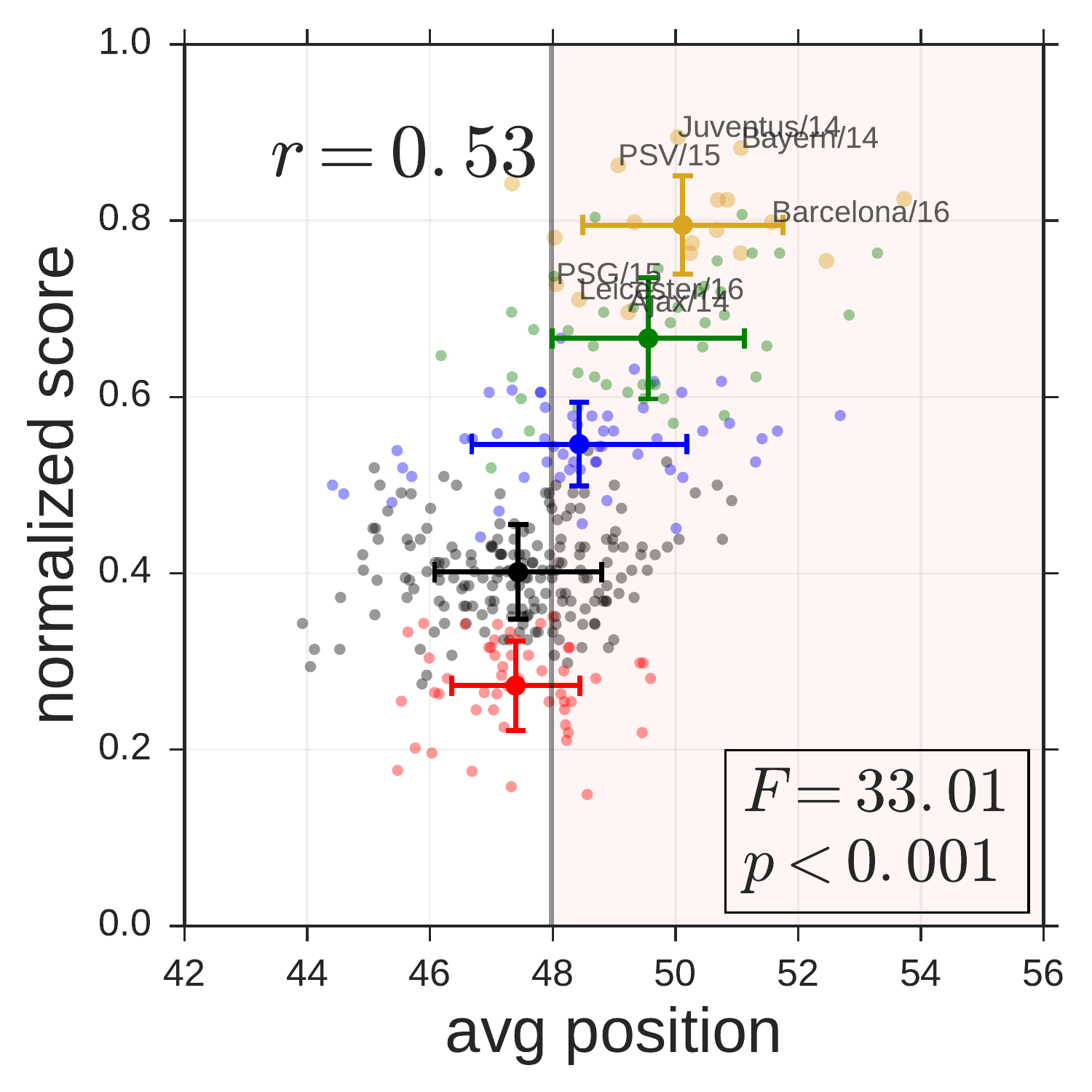}}
\subfigure[]{\includegraphics[scale=0.27]{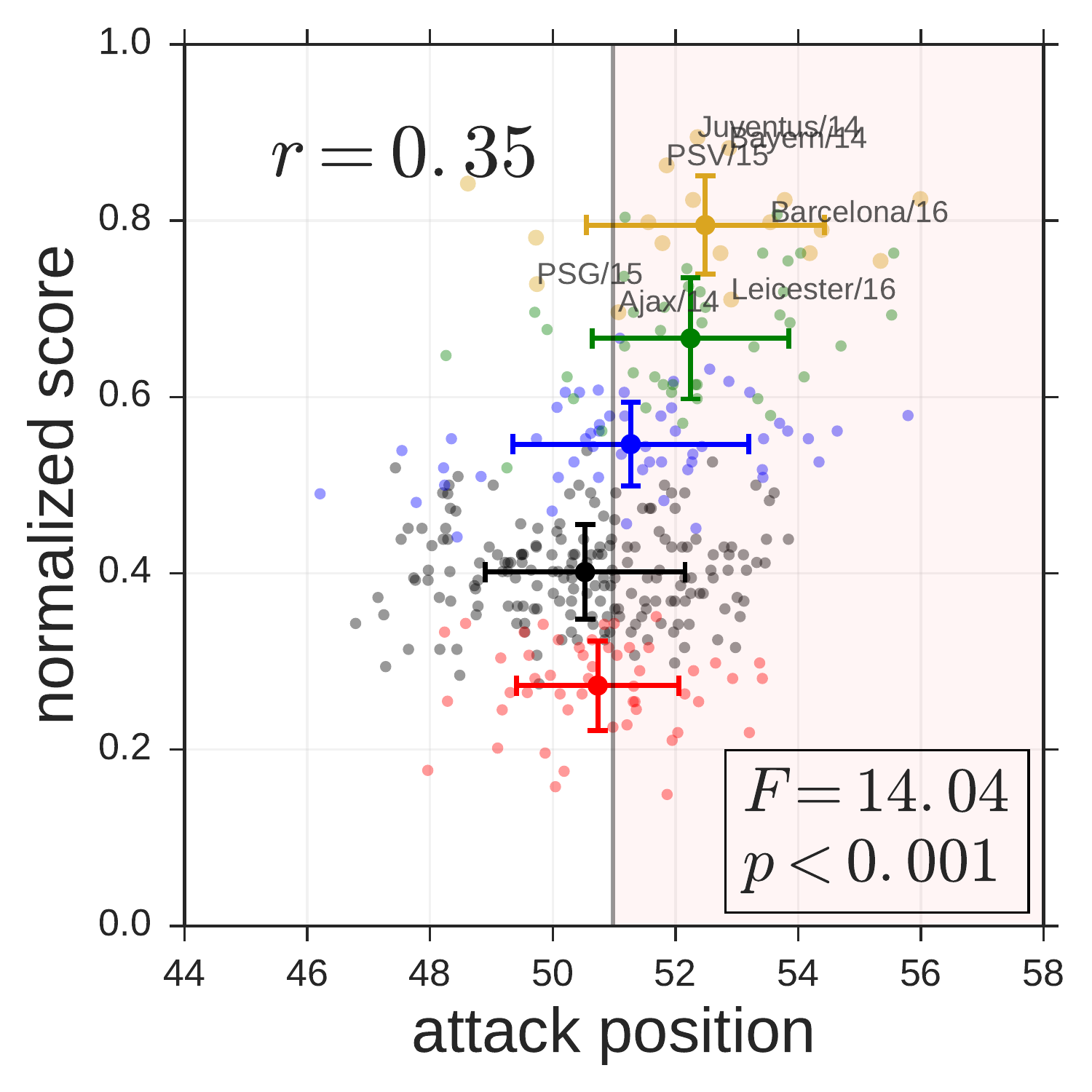}}
\subfigure[]{\includegraphics[scale=0.27]{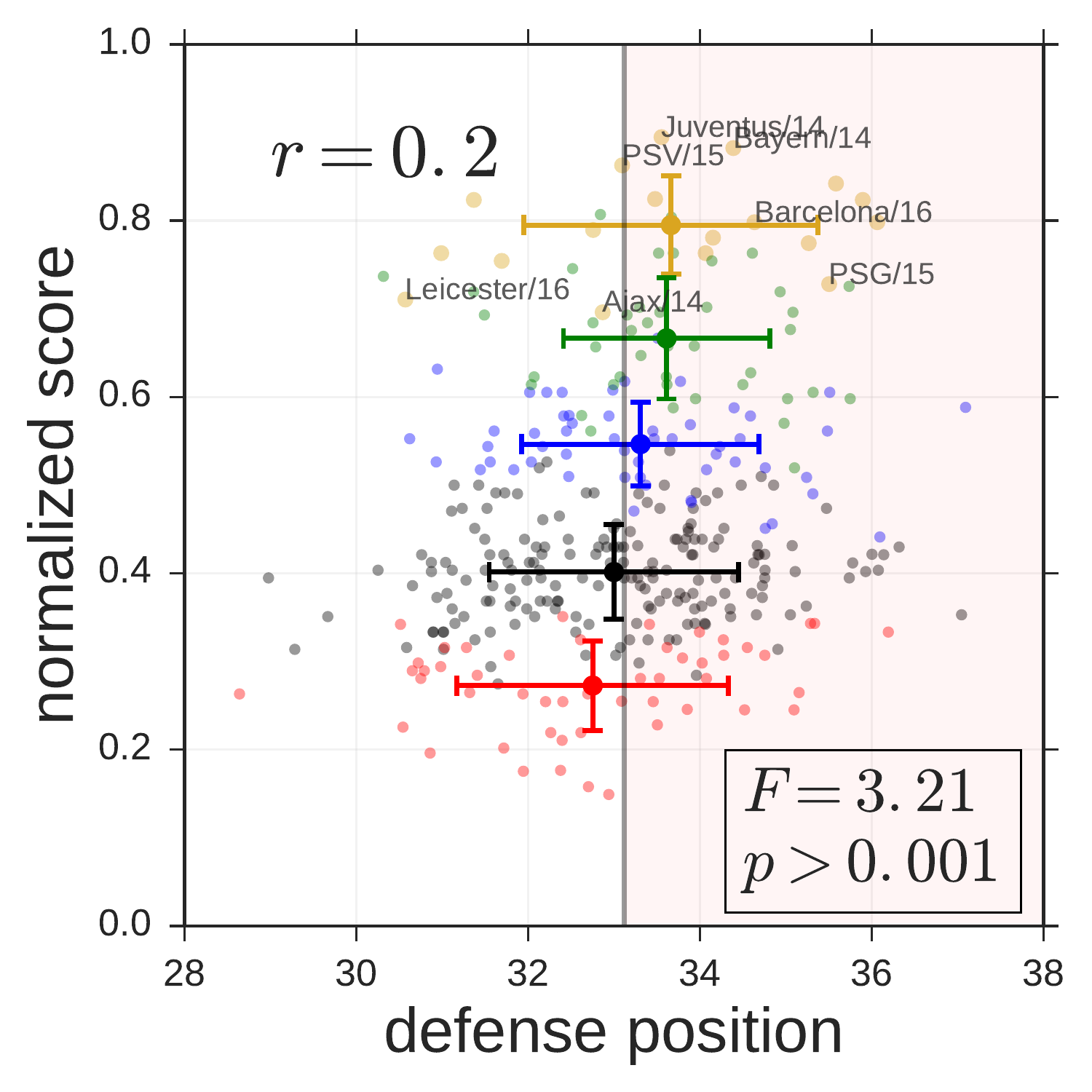}}

%second row
\subfigure[]{\includegraphics[scale=0.27]{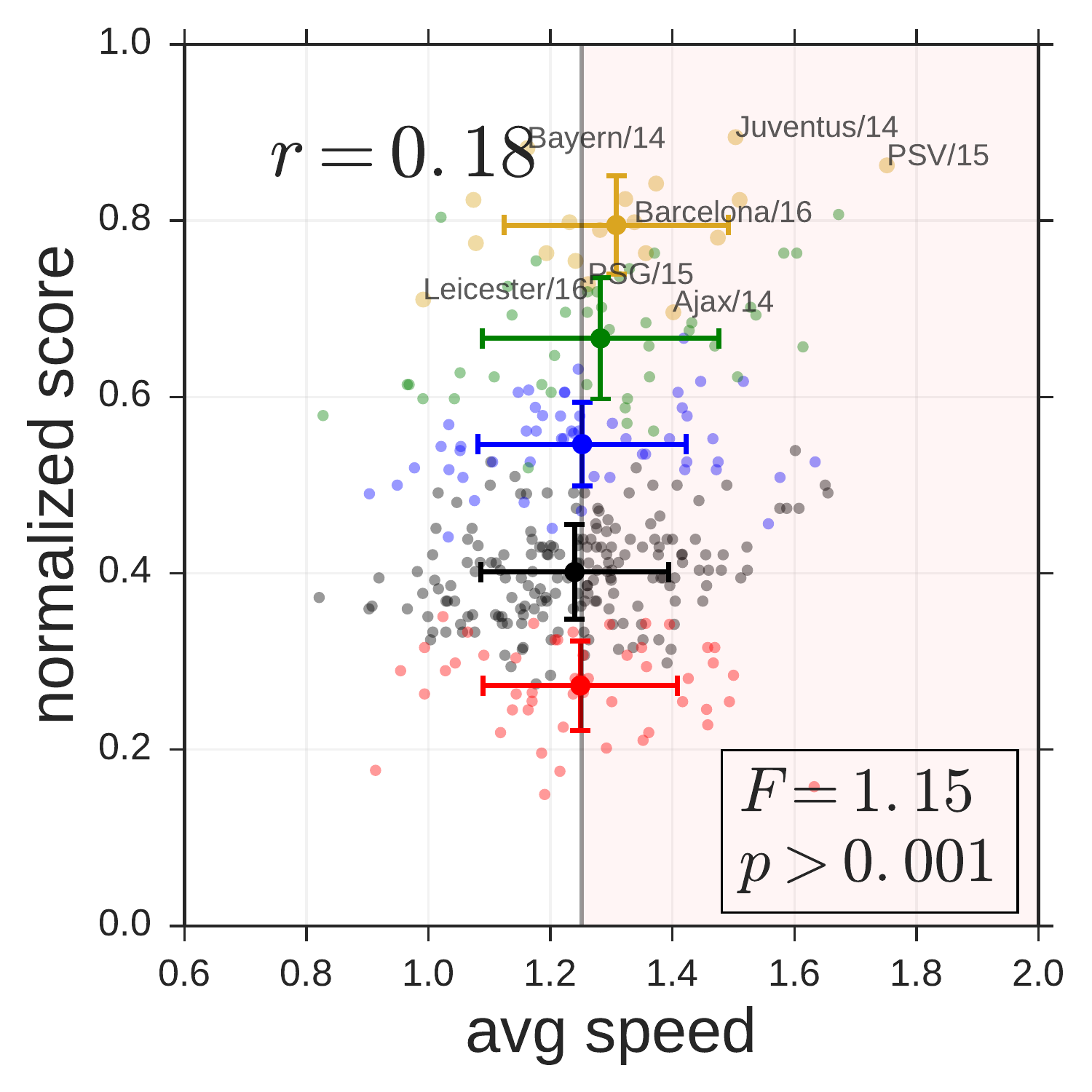}}
\subfigure[]{\includegraphics[scale=0.27]{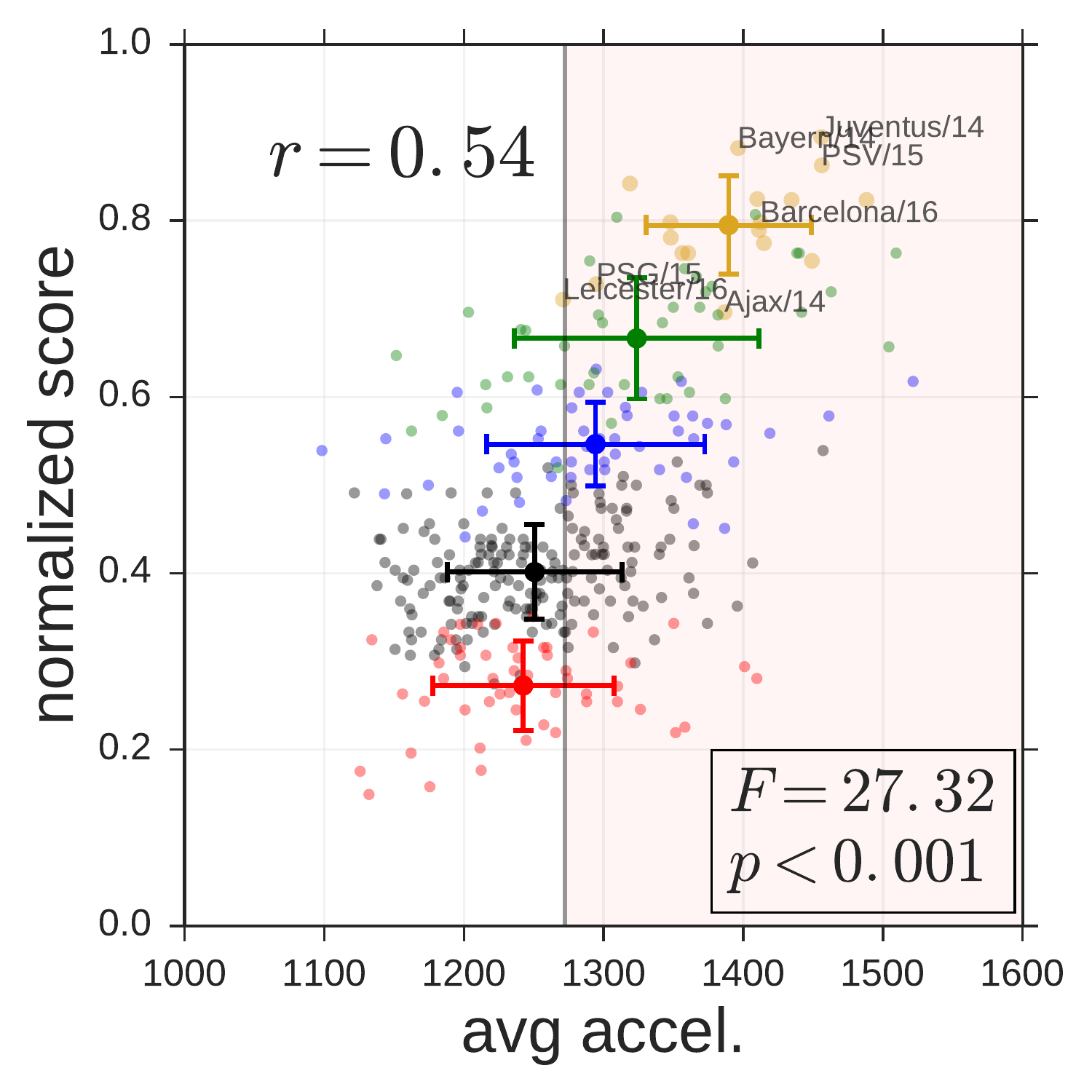}}
\subfigure[]{\includegraphics[scale=0.27]{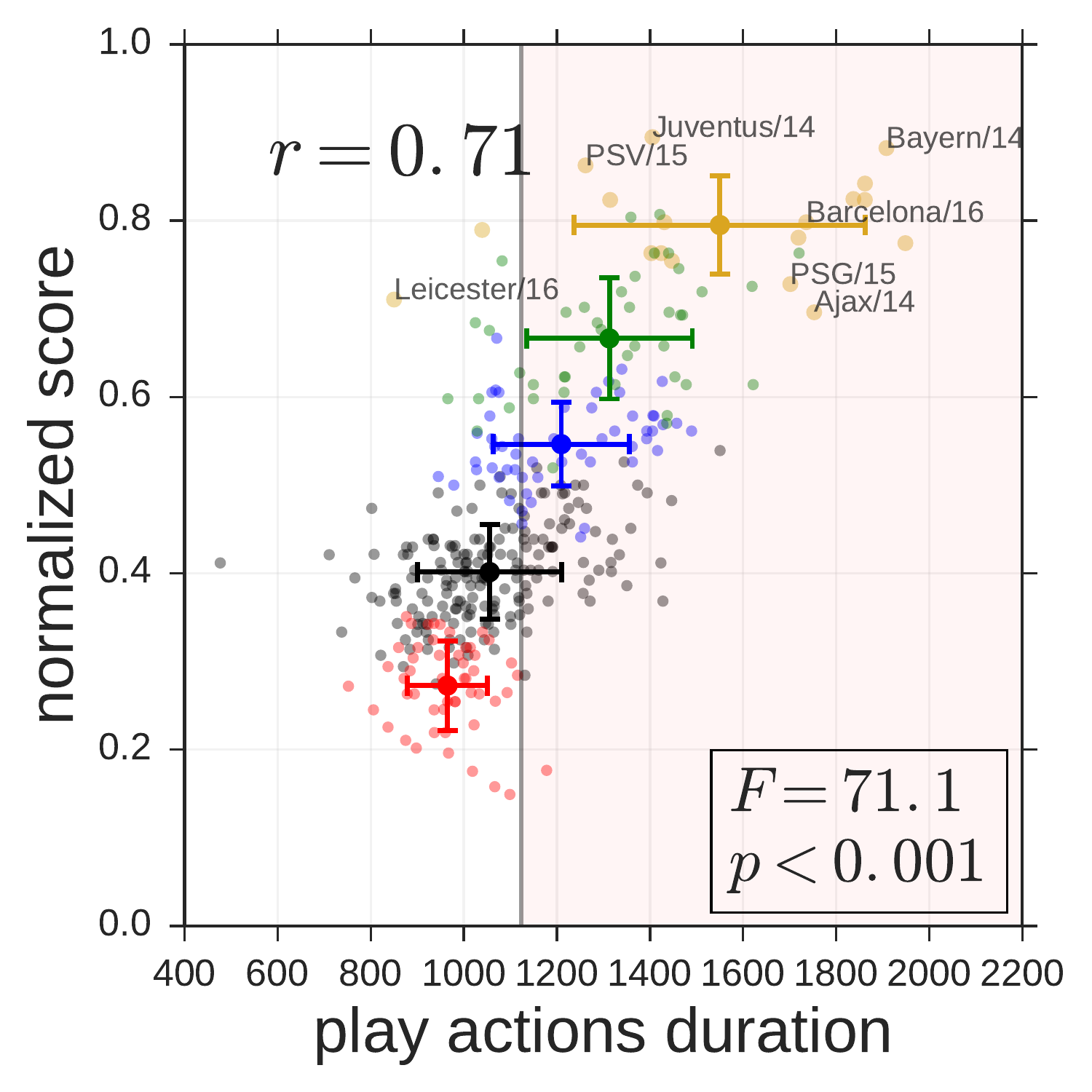}}

\caption{\textbf{Correlation between spatio-temporal features and success}. We split the teams in groups according to the final ranking: teams relegated in second division (red), teams in the middle (black), teams in Europa League (blue), teams in Champions League (green) and winners (golden). In the box we indicate F-test statistic and p-value resulting from a one-way ANOVA to assess whether the values of features within the groups differ from each other.}
\label{fig:new_corrs_dominance}
\end{figure}

While tackle precision and cross precision have null correlations with success (see Figure \ref{fig:new_corrs}c and \ref{fig:new_corrs}e), pass precision ($r=0.64$), dribble precision ($r=0.23$), corners ($r=0.57$) and attack/defense ratio ($r=0.68$) are strongly correlated with success (see Figure \ref{fig:new_corrs}a-b, \ref{fig:new_corrs}d-f). Regarding the dominance features, they are all correlated with success (Figure \ref{fig:new_corrs_dominance}). Nevertheless, the OLS process fitted on the extended set of features produces results comparable to the OLS fitted on the initial feature set, both for the absolute performance case ($R^2 = 0.59$) and the relative performance case ($R^2 = 0.67$). These results suggest that, while the quality and dominance features are individually correlated with success, they do not add significant predictive power in the OLS process, presumably because the information they represent is already subsumed by the combination of the ten considered features. 

\section{Normalization of final scores}
\label{sec:norm_scores}
Two clubs in different leagues may play a different number of games. For example, each Italian club plays 38 games during the season, while each German club plays 34 games. The total number of points that a club can gain during the season hence varies from league to league. To make the number of points comparable, we normalize it in the range $[0, 1]$. We use the following normalization function $norm(x) = \frac{x}{max(n)}$, where $x$ is the number of points gained by the team in the final ranking of the season, $max(n) = 3*2*(n-1)$ is the maximum number of points a club can gain during a season in a league with $n$ clubs. To clarify this concept, let us consider two clubs $T_1$ and $T_2$ which both gain 100 points in the corresponding leagues $L_1$ and $L_2$, where in $L_1$ clubs play 38 games and in $L_2$ clubs play 34 games. After normalization, team $T_1$'s score is 0.87 points while $T_2$'s score is 0.98, indicating that team $T_2$ gained almost all the points available during the season.

\section{Hyper-parameters tuning and cross-validation}
\label{app:cross_validation}
We performed our experiments through a rigorous validation process, in order to avoid issues related to hypothesis testing and overfitting. In particular, we perform hyper-parameter optimization to find the best combination of a machine learning model's hyper-parameters (i.e., the combination of parameters leading to the best predictive results). However, hyper-parameter optimization can lead to overfitting: if we run experiments on the same train-test splits, then performance on the test data is being incorporated into the training data by the choice of hyper-parameters. 

To avoid this problem, for every prediction experiment we split the dataset $D$ into two parts. The first part, $V$, accounting for 20\% of the dataset, is used to tune the model hyper-parameters.\footnote{We use the Python package \texttt{scikit-learn} and object GridSearchCV for hyper-parameters tuning (\url{http://bit.ly/2sIJSXK}). GridSearchCV trains a model for each combination of the parameters' values using cross-validation and outputs the combination leading to the best predictive results.} The remaining 80\% of data, $C$, is used to perform a 10-folds cross-validation \cite{friedman2001elements}. Cross-validation is a model validation technique for assessing how the results of a statistical analysis will generalize to an independent dataset. One of the main reasons for using cross-validation is that the conventional holdout validation (e.g., partitioning the dataset into two sets of 70\% for training and 30\% for test) can lead to a significant loss of modelling or testing capability. In 10-folds cross-validation, the dataset $C$ is split into ten parts: in turn, nine parts are used for training the model and the remaining part as a test set for validation. Both the hyper-parameters tuning on a separated portion of the dataset and the cross-validation strategy avoid the risk of overfitting and guarantee that predictive results are significant and reliable. 

\section{Post hoc Tukey's test}
\label{app:anova}

\begin{figure}[htb]\centering
\subfigure[]{\includegraphics[scale=0.4]{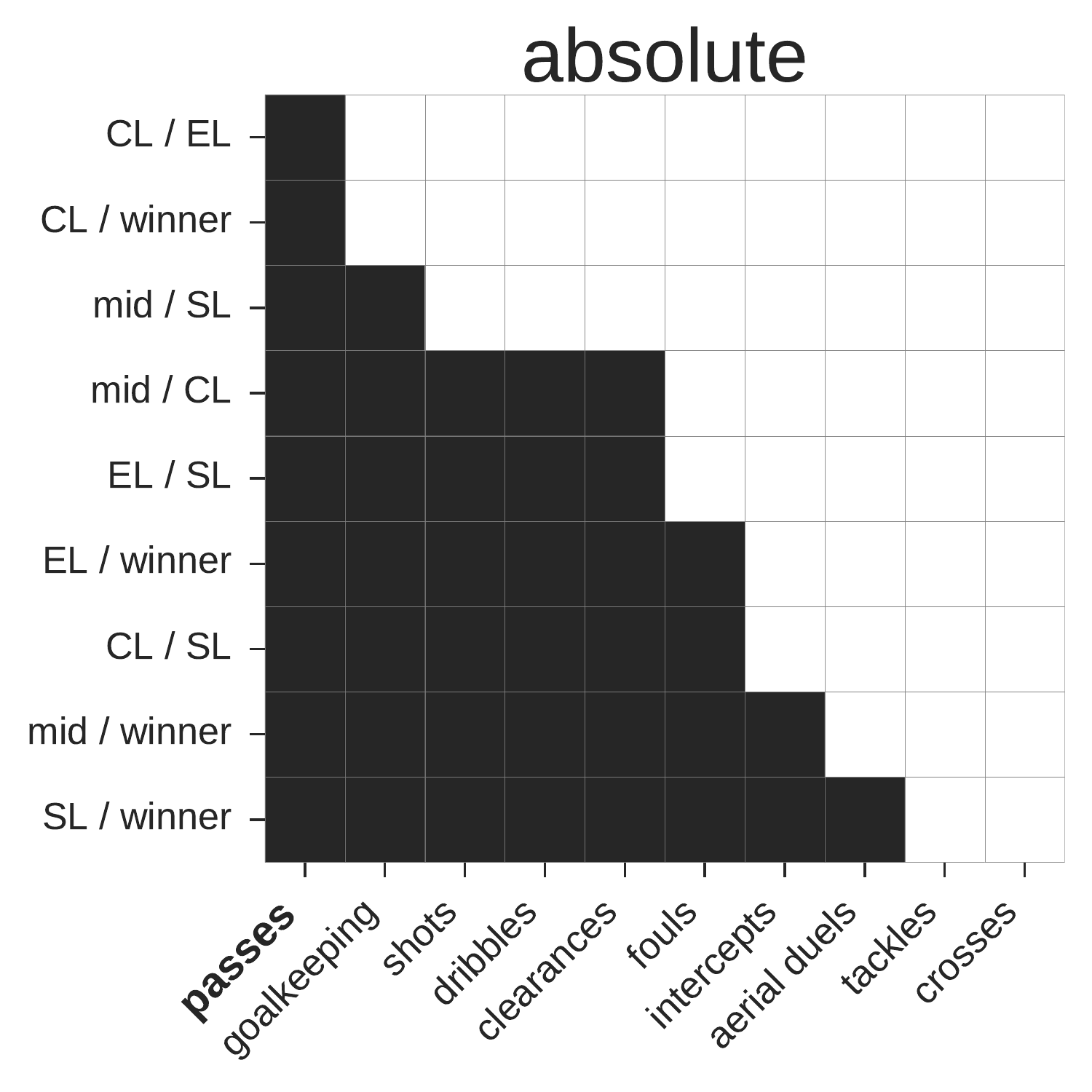}}
\subfigure[]{\includegraphics[scale=0.4]{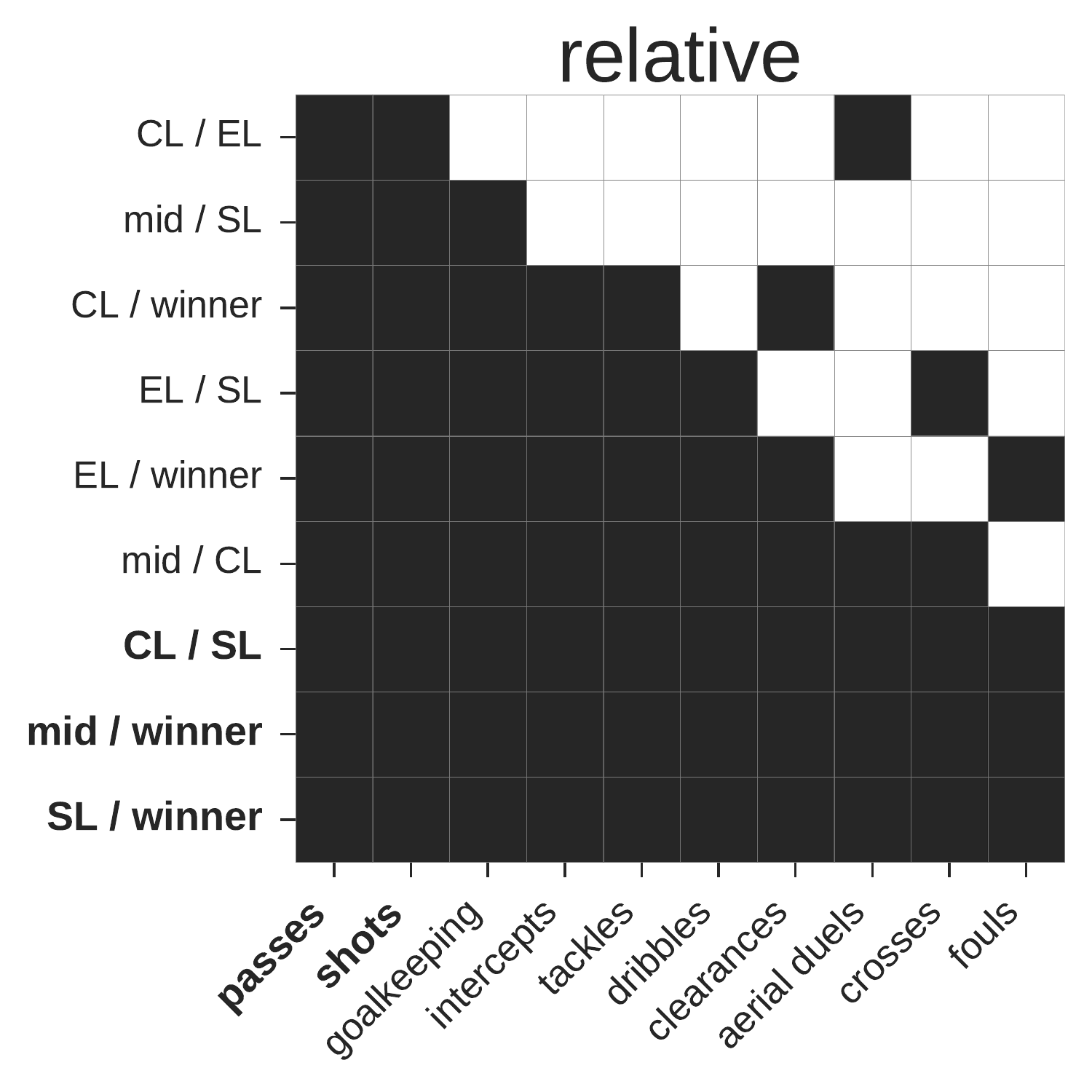}}
\caption{\textbf{Results of Tukey's test.} Heatmap representing a matrix indicating, for every pair of groups of success and absolute performance feature, whether (black) or not (white) the difference between the means of the two groups are significant.}
\label{fig:tukey}
\end{figure}

Tukey's range test \cite{tukey} is a common method used as post hoc analysis after one-way ANOVA. We perform this test to address the multiple comparison's
problem. Indeed, as many technical features are compared, it becomes increasingly likely that the success groups will appear to differ on at least one feature due to random sampling error alone.
The Turkey's range test compares all possible pairs of groups to identify difference between two means greater than the expected standard error. We use the \texttt{statsmodels} package
in Python to perform Tukey's range test. Figure \ref{fig:tukey}a shows a matrix indicating, for each success group and feature pair, whether (black) or not (white) the difference 
between the means of the two groups is significant.

\section{Classification experiments}
\label{app:classification}
We evaluate the goodness of classification with a 10-fold cross-validation strategy by two metrics: \emph{(i)} the accuracy of classification $\mbox{ACC} {=} {|\hat{f}(\overline{\vect{h}}_i) = f(\overline{\vect{h}}_i)| \over n}$, where $f(\overline{\vect{h}}_i)$ is the actual class of success of team $i$, $\hat{f}(\overline{\vect{h}}_i)$ is the predicted class, and $n$ is the number of teams in the training dataset \cite{tan_introduction_2005}; \emph{(ii)} the weighted average f-score, defined as $F1 = \sum_{c \in C} |c| {2\mbox{\small TP} \over {2\mbox{\small TP} + \mbox{\small FP} + \mbox{\small FN}}}$, where TP, FP, FN are the numbers of true positives, false positives and false negatives resulting from classification, $C{=}\{$\texttt{top}, \texttt{bottom}$\}$ is the set of the two classes of success and $|c|$ is the support of a class \cite{tan_introduction_2005}. 

\begin{figure}[H]\centering
\subfigure[]{\includegraphics[scale=0.4]{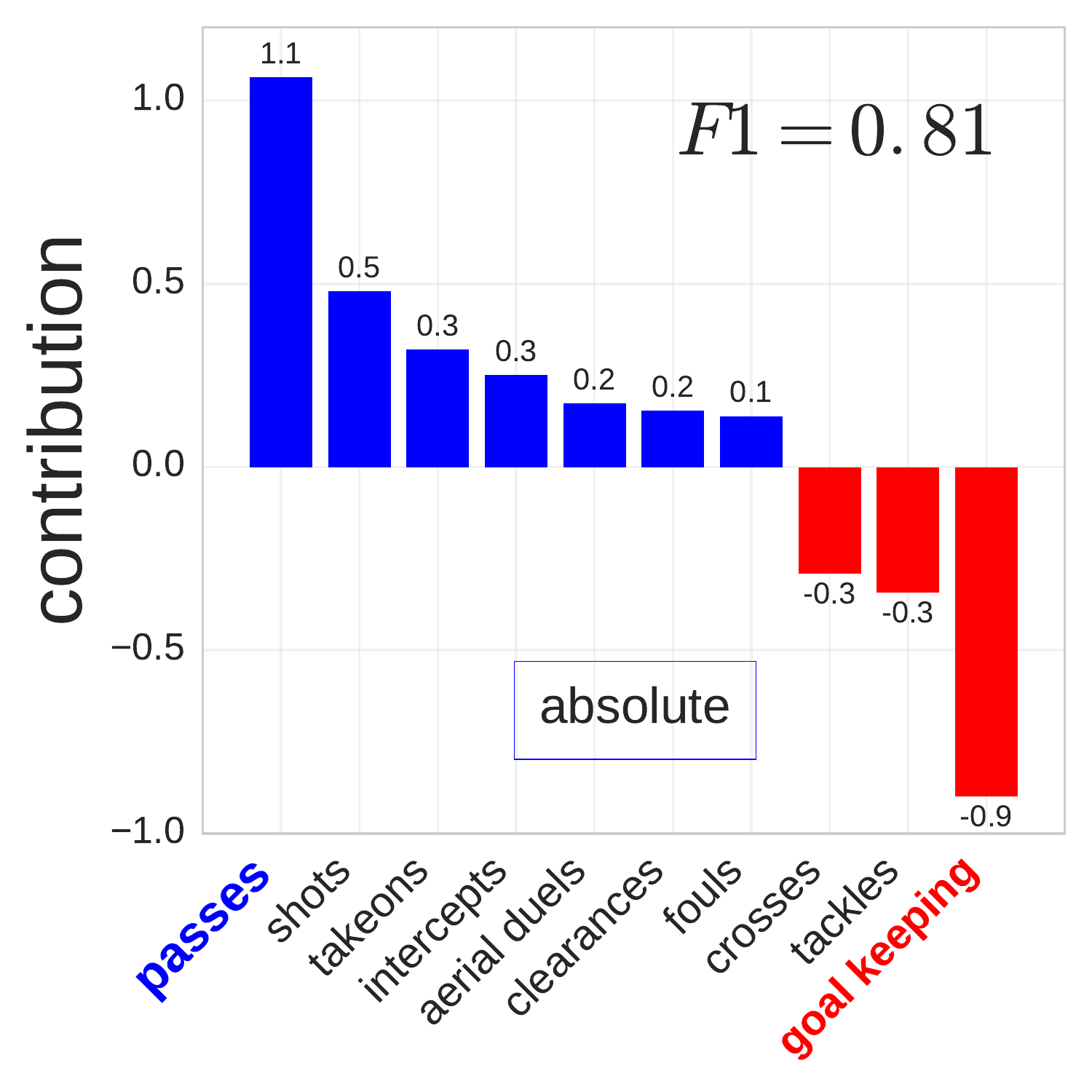}}
\subfigure[]{\includegraphics[scale=0.4]{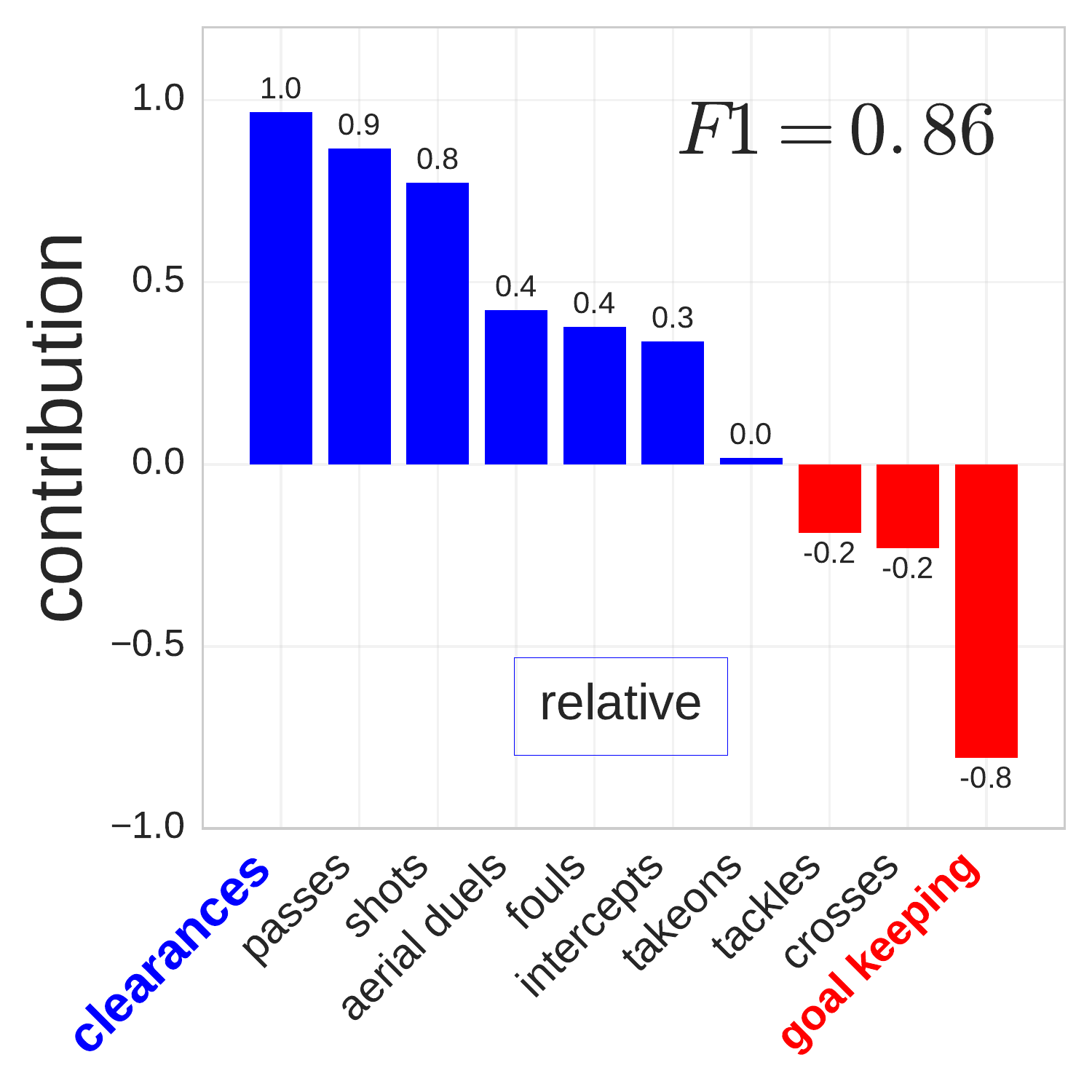}}
\subfigure[]{\includegraphics[scale=0.4]{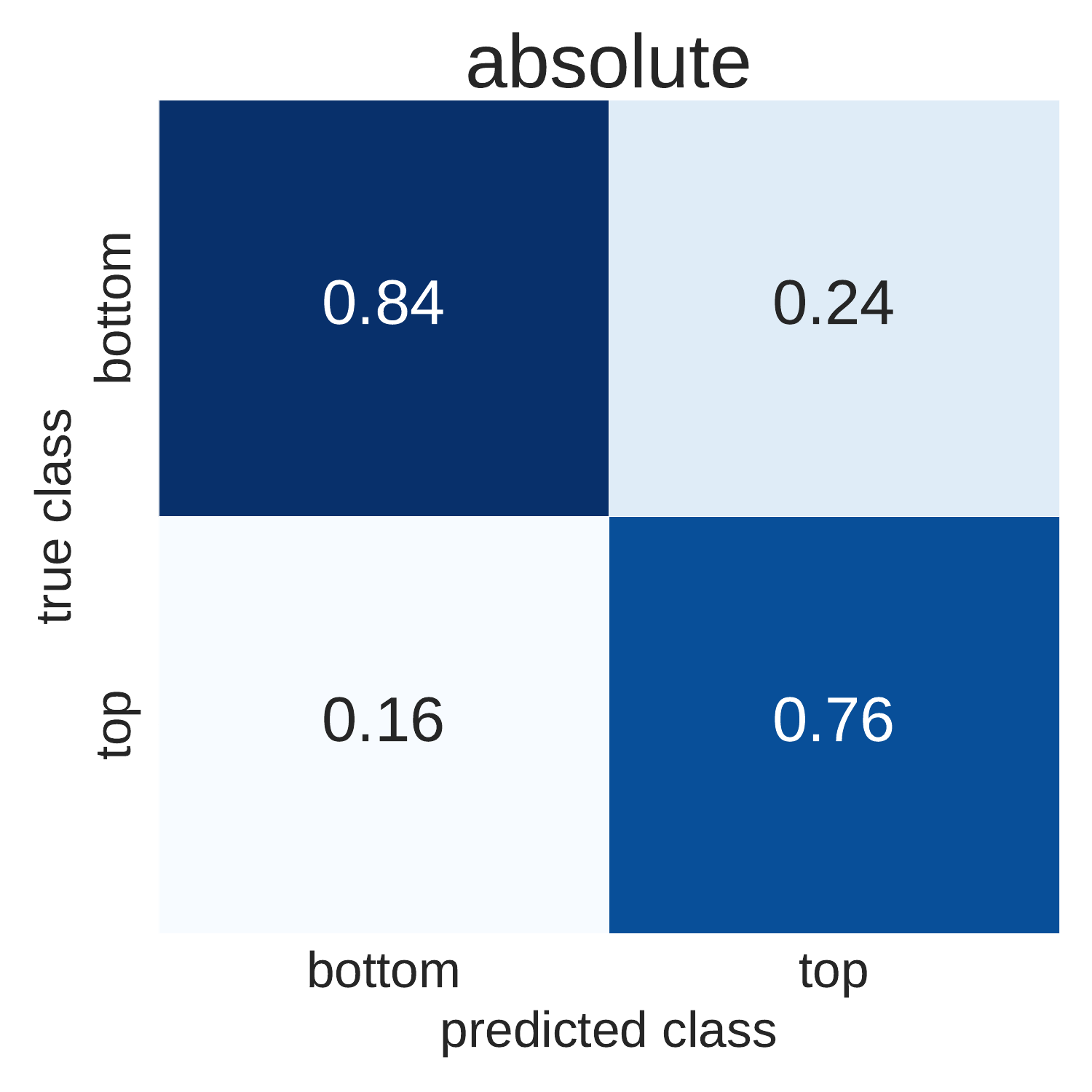}}
\subfigure[]{\includegraphics[scale=0.4]{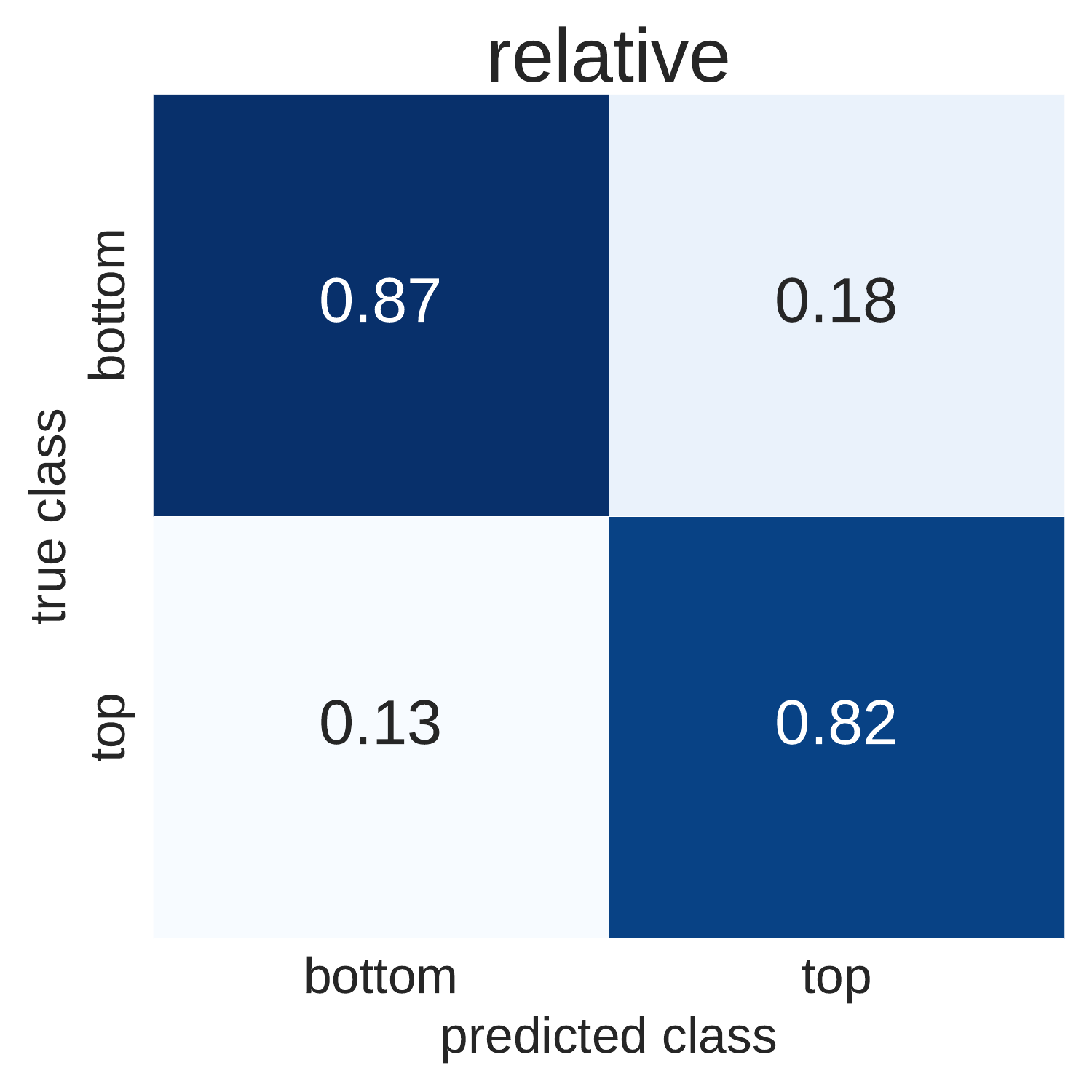}}

\caption{\textbf{Results of the classification of teams into groups of success.} Coefficients produced by the logistic regression for absolute performance features (a) and relative performance features (b). Classification matrices for absolute (c) and relative (d) performance features.}
\label{fig:class_importance}
\end{figure}

For the absolute performance features we train a logit $C_{abs}$ and observe ACC=0.82 and $F1{=}0.81$, significantly better than a baseline classifier which always predicts the most frequent class (\texttt{bottom}) which has ACC=0.67 and $F1{=}0.54$. (Table \ref{tab:pred_results}). Similarly to the regression case, a classification model $C_{rel}$ trained on the relative performance features produces better results than the absolute case, with ACC=0.86 and $F1{=}0.86$. These results show that multidimensional performance can discriminate between top teams and bottom teams.
Figure \ref{fig:class_importance} shows a matrix representing the classification error for each class of success, for $C_{abs}$ (c) and $C_{rel}$ (d). An element $i, j$ in the matrix indicates the fraction of instances for which a team in class $j$ is classified as a team in class $i$ by the logit. The diagonal of the matrix, hence, indicates the classifier's recall for every label, i.e., the fraction of teams in the class that are correctly classified by the logit. We find that the recall of both the \texttt{top} and the \texttt{bottom} classes are high, indicating that the classifier accurately discriminates between the two classes of success. 

Figure \ref{fig:class_importance}a-b shows the importance of each feature to the classification task. For absolute performance, the results agree with the regression task: a team's typical number of passes is the strongest predictor of its level of success; followed by the typical number of goalkeeping actions and the typical number of shots. For relative performance, the typical difference in clearances is the strongest predictor of success, followed by the typical difference in ball possession and goalkeeping actions. 

\section{Differences between the actual and the PC ranking}
\label{sec:simulation_results}

\begin{figure}[H]\centering
\includegraphics[scale=0.35]{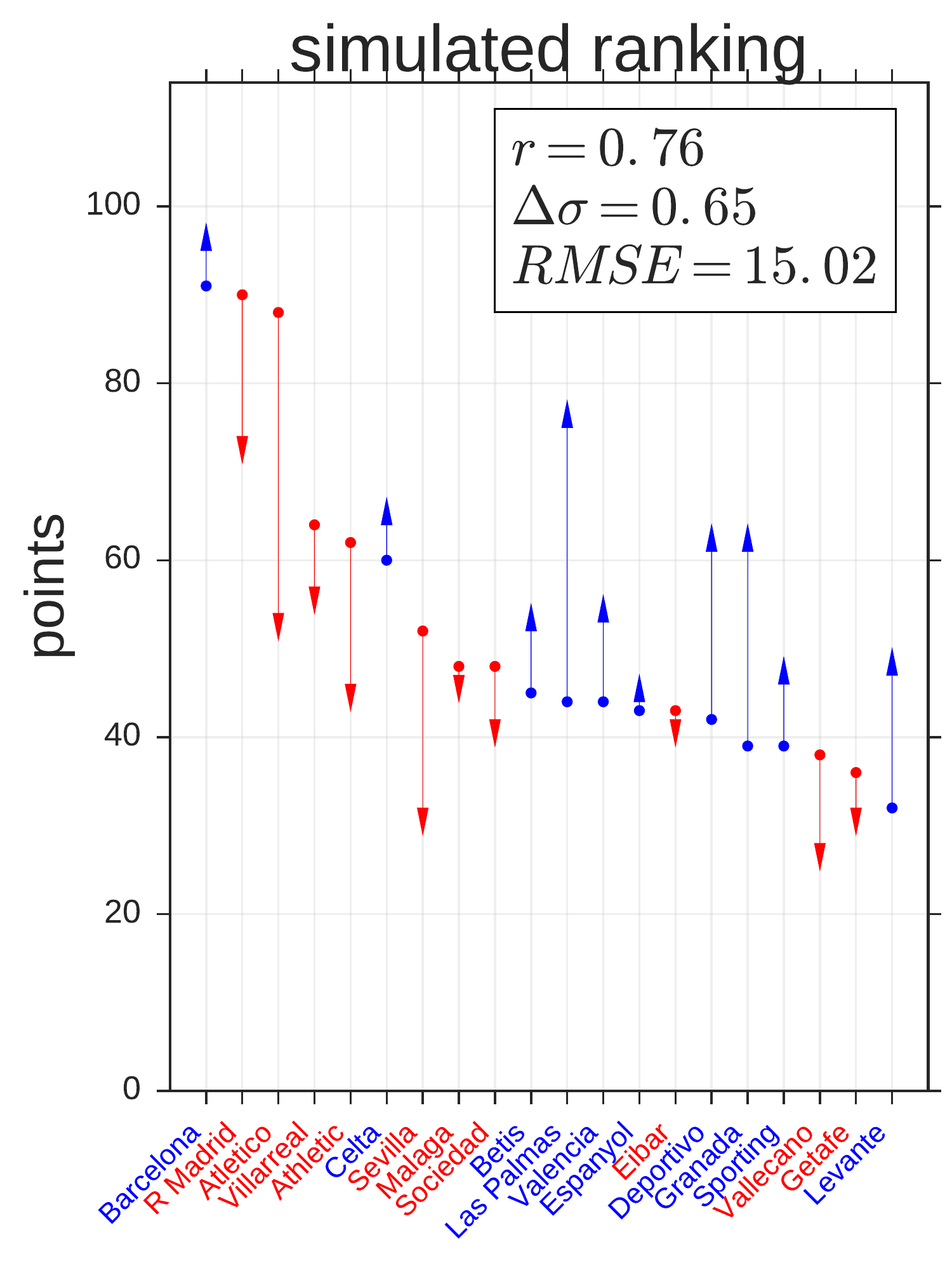}
\caption{\textbf{Errors between actual ranking and PC rankings in La Liga.} The dots indicate the points in the actual ranking; the length of the arrows indicate the error of simulation. The blue arrows indicate that the simulation overestimates the number of points with respect to the actual ranking, the red arrows indicate that the simulation underestimates the number of points.}
\label{fig:errors}
\end{figure}

Figure \ref{fig:errors} visualizes the difference between the actual ranking and the PC ranking for La Liga 2015/2016. Our simulation tends to underestimate the number of points for the teams in the top of the ranking and to overestimate the number of points for the teams in the bottom of the ranking (Figure \ref{fig:errors}). 

\section*{Acknowledgments}
We thank Alessio Rossi and Marco De Nadai for their useful suggestions. This work has been partially funded by the EU project SoBigData grant n. 654024.

\bibliographystyle{ws-acs}
\bibliography{biblio.bib}  

\end{document}